\pdfoutput=1

\documentclass[10pt]{iopart}	
\usepackage{graphicx}
\usepackage{iopams}
\usepackage{cite}
\usepackage{xcolor}

\usepackage{academicons}
\usepackage{braket}
\usepackage{subcaption}		

\usepackage{placeins}				
\usepackage[normalem]{ulem}  		
\usepackage{cancel}					

	\newcommand{\Orf}[0]{\Omega_\mathrm{rf}}
	\newcommand{\Orfind}[1]{\Omega_{\mathrm{rf},#1}}
	\newcommand{\osec}[0]{\omega_\mathrm{sec}}
	\newcommand{\osecind}[1]{\omega_{\mathrm{sec},#1}}
	\newcommand{\ocom}[0]{\omega_\mathrm{com}}
	
	\newcommand{\orcid}[1]{\href{https://orcid.org/#1}{\textcolor[HTML]{A6CE39}{\aiOrcid}}}

\begin{document}


\title{Motional heating of spatially extended ion crystals}

	\author{D~Kalincev$^1$, LS~Dreissen$^1$, AP~Kulosa$^1$, C-H~Yeh$^1$, HA~F\"urst$^{1,2}$, TE~Mehlst\"aubler$^{1,2}$}
\address{$^1$ Physikalisch-Technische Bundesanstalt (PTB), Bundesallee 100, 38116 Braunschweig, Germany}
\address{$^2$ Institut f\"ur Quantenoptik, Leibniz Universit\"at Hannover, Welfengarten 1, 30167 Hannover, Germany}

\ead{tanja.mehlstaeubler@ptb.de}
\date{\today}

\begin{abstract}
We study heating of motional modes of a single ion and of extended ion crystals trapped in a linear radio frequency (rf) Paul trap with a precision of $\Delta \dot{\bar{n}} \approx 0.2$\, phonons\,s$^{-1}$. Single-ion axial and radial heating rates are consistent and electric field noise has been stable over the course of four years. At a secular frequency of $\osec=2\pi\times620$\,kHz, we measure $\dot{\bar{n}} = 0.56(6)$\,phonons\,s$^{-1}$ per ion for the center-of-mass (com) mode of linear chains of up to eleven ions and observe no significant heating of the out-of-phase (oop) modes. By displacing the ions away from the nodal line, inducing excess micromotion, rf noise heats the com mode quadratically as a function of radial displacement $r$ by $\dot{\bar{n}}(r)/r^2 = 0.89(4)$\,phonons\,s$^{-1}$\,$\mu$m$^{-2}$ per ion, while the oop modes are protected from rf-noise induced heating in linear chains. By changing the quality factor of the resonant rf circuit from $Q=542$ to $Q=204$, we observe an increase of rf noise by a factor of up to 3. We show that the rf-noise induced heating of motional modes of extended crystals also depends on the symmetry of the crystal and of the mode itself. As an example, we consider several 2D and 3D crystal configurations. Heating rates of up to 500\,ph\,s$^{-1}$ are observed for individual modes, giving rise to a total kinetic energy increase and thus a fractional time dilation shift of up to $-0.3\times 10^{-18}$\,s$^{-1}$ of the total system. In addition, we detail how the excitation probability of the individual ions is reduced and decoherence is increased due to the Debye-Waller effect.
\end{abstract}

\noindent{\it precision metrology, ion Coulomb crystals, vibrational mode heating, rf noise, multi-ion clocks \/}\\ 

\section{Introduction}\label{intro}
Single-ion spectroscopy has lead to accurate optical atomic clock operation and enabled searches for new physics with high sensitivity over the past decades \cite{Safronova:2018}. 
This includes the search for a variation of fundamental constants \cite{Rosenband:2008,Huntemann:2014,Godun:2014}, dark matter \cite{Derevianko:2014}, a hypothetical fifth force \cite{Delaunay:2017,Flambaum:2018,Berengut:2018,Knollmann:2019,Solaro:2020,Counts:2020,Berengut:2020} and, in general, more precise tests of Einstein's theory of general relativity \cite{Dzuba:2016,Shaniv:2018,Sanner:2019,Lange:2020}.

In many cases, the resolution is limited by the poor signal-to-noise ratio from a single atomic absorber. To overcome this limitation, several new approaches have been proposed, which utilize multiple ions in a so-called Coulomb crystal, e.g.~based on Ca$^+$\cite{Champenois:2010}, Lu$^+$\cite{Arnold:2015} or In$^+$\cite{Herschbach:2012,Keller:2019b} ions trapped in linear radio frequency (rf) Paul traps. Also, quantum simulations with trapped ions is advancing towards 2D and 3D systems to simulate more complex Hamiltonians \cite{Zhang:2017,Richerme:2016}. However, new challenges arise from extending the crystal size and dimension. These are both of fundamental nature to trapping ions in rf-traps, e.g.~frequency shifts induced by excess micromotion (EMM) or additional electric field gradients from neighbouring ions, and of a technical nature, e.g.~reaching the required homogeneity of spectroscopy beam intensities, electric and magnetic fields over a larger region. Many of these effects have been investigated in linear ion chains, where micromotion can be well controlled and uncertainties on frequency shifts are expected to be at the $10^{-19}$ level \cite{Keller:2019b,Tan:2019}. For two- and three-dimensional crystals, where high rf electric fields are probed, techniques for cancelling rf field shifts have been investigated \cite{Berkeland:1998,Barrett:2015,Dube:2014,Huang:2016}.

In this article, we take a linear chip trap with well controlled electromagnetic fields as a platform to investigate the heating of ion Coulomb crystals in two scenarios, i.e.~under influence of noise induced by static electric fields (dc) and noise induced by rf fields. From this, we extrapolate its impact onto 2D and 3D crystals. We measure dc heating of the centre-of-mass (com) mode and higher modes in a linear chain of ions with an uncertainty of $\sim$0.1\,phonons s$^{-1}$ and find that the dc heating along the radial direction has been consistent over the course of four years.

By applying controlled radial EMM we can systematically amplify the coupling of rf noise to the ion motion, leading to an enhanced heating rate of the com mode. By filtering the rf noise by resonant circuits of different quality factors, we deduce the power spectral density of the electric field noise. Using the experimentally obtained parameters, we investigate heating of larger radially extended 2D and 3D ion crystals and find that rf-noise induced heating of modes strongly depends on the symmetry of the crystal and of the modes themselves. We also calculate the time dilation shifts and the influence from the Debye-Waller effect on each individual ion for spectroscopy applications of extended crystals.
\section{Experimental setup and methods}\label{experiment}
\subsection{Experiment}
\begin{figure}[t]
	\begin{center}
		\includegraphics[width=0.8\textwidth]{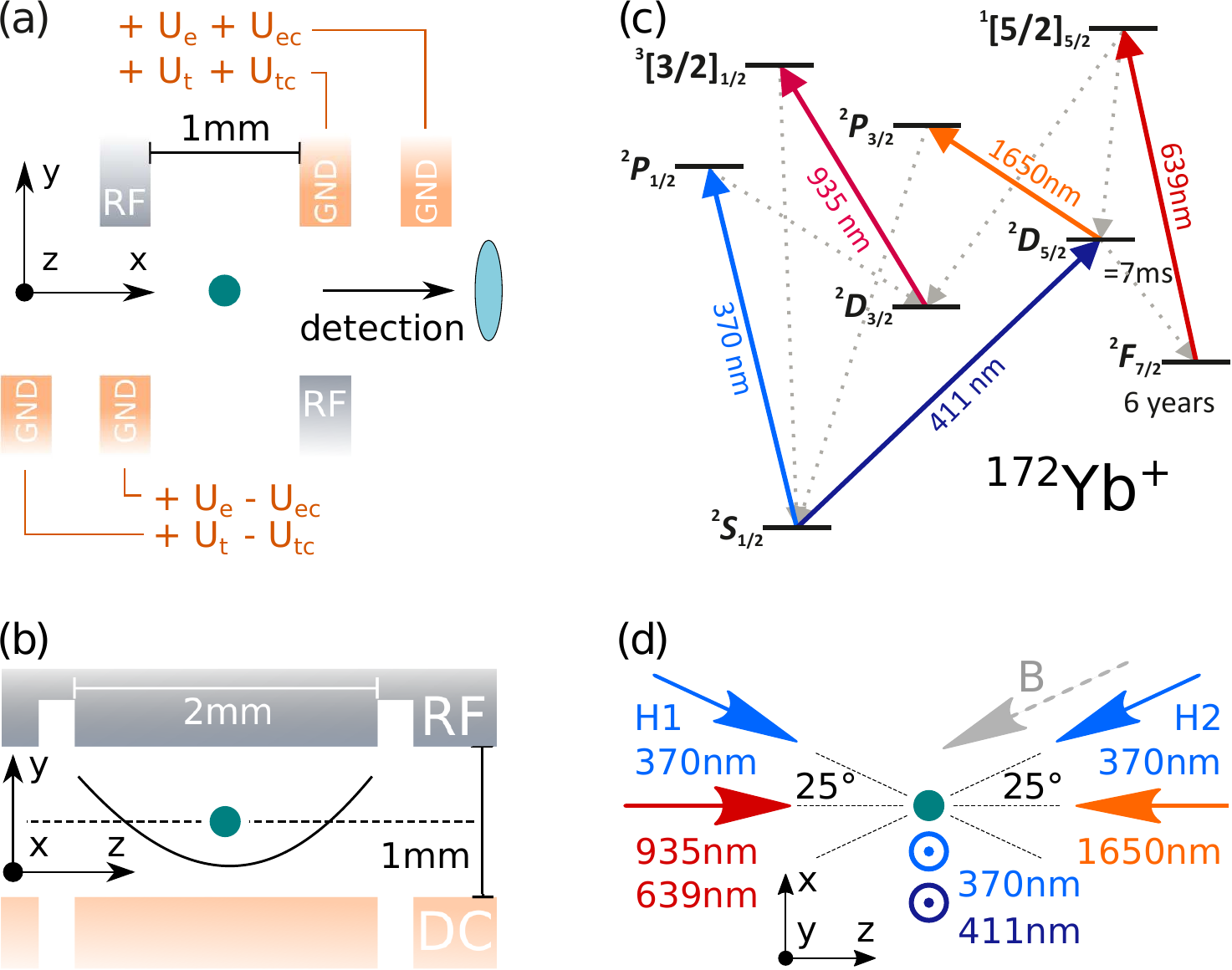}
	\end{center}
		\caption{Experimental setup and the energy level scheme of $^{172}$Yb$^+$. }{
	Ion trap geometry is shown in the (a) $xy$ plane and (b) in the $zy$ plane. The ions are trapped in a 2\,mm segment of the linear Paul trap, where the adjacent segments provide axial confinement.  Additional voltages may be applied to rotate the radial axes ($\mathrm{U_t}$, $\mathrm{U_e}$) and to compensate stray electric fields ($\mathrm{U_{tc}}$, $\mathrm{U_{ec}}$). \\
(c) Reduced energy level scheme of the $^{172}$Yb$^+$ ion, showing the relevant transitions and decay channels for Doppler cooling (370\,nm), sideband cooling (411\,nm and 1650\,nm), repumping  (935\,nm and 639\,nm) and spectroscopy (411\,nm). \\
(d) The repump lasers are aligned along the trap axis $z$, while the spectroscopy laser at 411\,nm and one of the radial Doppler cooling lasers at 370\,nm are aligned vertically along $y$. Two 370\,nm lasers in the $xz$ plane are used for micromotion measurements (H1 and H2) and for state preparation through optical pumping (H2). The quantization axis is set by the magnetic field ($B$), which is parallel to H2.}
	\label{fig:1}
\end{figure}

The experiment is based on sideband thermometry of ground-state cooled $^{172}$Yb$^+$ ions, which are stored in a linear rf Paul trap. The trap consists of four wafers with segmented electrodes (\fref{fig:1}(a) and (b)). The radial confinement is set by an rf electric field supplied by a resonant circuit to the inner two wafers, where additionally a combination of DC voltages may be applied to lift the degeneracy and to rotate the principal axes of the radial confinement ($\mathrm{U_t}$, $\mathrm{U_e}$) and to compensate for stray electric fields ($\mathrm{U_{tc}}$, $\mathrm{U_{ec}}$). The axial confinement in the trapping segment is provided by the voltages $\mathrm{U_t}$ of the neighbouring segments. The trap rf drive frequency is $\Orf = 2\pi \times 24.4$\,MHz and typical secular frequencies are $ \omega_{x,y,z} = 2\pi \times (600,580,205)$\,kHz. The trap was designed for low axial EMM and was proposed as a platform for scalable multi-ion clock-based experiments with mixed species of $^{115}$In$^+$ and $^{172}$Yb$^+$ \cite{Herschbach:2012,Keller:2019b,Pyka:2014}.
A detailed description and characterization of the Paul trap and the laser systems can be found in \cite{Burgermeister:2019} and \cite{Keller:2015}, respectively.
\Fref{fig:1}(c) and (d) show the relevant part of the atomic energy level scheme of $^{172}$Yb$^+$ and the laser geometry, respectively. The ions are loaded from a thermal atomic beam via two-step photo-ionization and stored by laser cooling on the $\mathrm{^2S_{1/2}}$$\rightarrow$$\mathrm{^2P_{1/2}}$ electric dipole transition at 370\,nm. The cooling-cycle is closed using a repump laser at  935\,nm. Fluorescence from the decay of the short lived $\mathrm{^2P_{1/2}}$-state is detected with an electron multiplying charged-coupled device (EMCCD) camera. At the beginning of the experimental sequence, a reference image is recorded with 2.5\,ms exposure time to decide whether the measurement cycle is valid, i.e.~whether the ion is in the cooling cycle or not. The ion is then cooled to the Doppler limit of about 0.5\,mK via a radial beam with a wavelength at  370\,nm along the $y$ direction. Optical pumping into the $\ket{\mathrm{^2S_{1/2}}},\mathrm{m_j}=-1/2$~Zeeman-substate is achieved via a $\sigma^-$-polarized beam at 370\,nm (H2 in \fref{fig:1}(d)), aligned along the quantization axis set by a magnetic field of~$B = 65\,\mu$T  at an angle of 25$^{\circ}$ with respect to the trap axis $z$. 

From here, the ion is cooled further to its quantum mechanical ground state via quench-assisted resolved sideband cooling (RSC) on the electric quadrupole transition $\mathrm{^2S_{1/2}}$$\rightarrow$$\mathrm{^2D_{5/2}}$ and the dipole allowed $\mathrm{^2D_{5/2}} \rightarrow \mathrm{^2P_{3/2}}$. Similar schemes have been previously reported for Ca$^+$ ions, see e.g.~\cite{Hendricks:2008,Lindenfelser:2017}. For a single ion, this continuous-wave (cw) sideband cooling scheme is applied for 5\,ms along the $y$ direction, the principal axis of the higher-frequency radial mode (within 3$^{\circ}$), to achieve $\bar{n} \leq 0.1$\,phonons (ph). Here, $\bar{n}$ describes the mean motional state occupation of an ion in the harmonic trap potential according to a thermal distribution.
For simultaneous sideband cooling of multiple motional modes of larger crystals, the frequency of the 411\,nm laser is tuned to the centre of the red sideband spectrum (again for the modes along the $y$ direction). The intensity of the 1650\,nm laser is first increased to sufficiently broaden the effective linewidth and address all motional modes simultaneously, trading final temperature for cooling time. In the final part of the cooling sequence, the intensity is reduced to cool the modes to $\bar{n} \leq 0.1$\,ph, within typical cooling times of 5\,-\,10\,ms. 
\subsection{Thermometry}\label{thermometry}
To determine the temperature of a certain mode, we use resolved sideband thermometry on the narrow electronic quadrupole transition \cite{Wineland:1998,Leibfried:2003}. This technique is based on detecting the ratio of the excitation probabilities of the first-order red ($p_\mathrm{r}$) and blue ($p_\mathrm{b}$) motional sidebands, from which we can determine the average motional state occupation according to \cite{Wineland:1998}
\begin{equation}
\label{eq:n}
\bar{n} = \frac{1}{p_\mathrm{b}/p_\mathrm{r}-1}.
\end{equation}
By varying the delay $\tau$ between the end of the sideband cooling cycle and the sideband interrogation pulse, we determine the heating rate. To detect the red and blue sideband amplitudes, a resonant first-order blue and red sideband pulse is alternately applied and the average excitation probability of each sideband is extracted from 200 measurement cycles.  
The uncertainty of a temperature measurement is given by
\begin{equation}
\label{eq:dn}
\Delta\bar{n} = \bar{n}^2\frac{p_\mathrm{b}}{p_\mathrm{r}} \sqrt{\bigg(\frac{\Delta p_\mathrm{b}}{p_\mathrm{b}}\bigg)^2+\bigg(\frac{\Delta p_\mathrm{r}}{p_\mathrm{r}}\bigg)^2}.
\end{equation}
The uncertainty of $\bar{n}$ increases at higher temperature, because it is directly proportional to $\bar{n}^2$. To obtain the highest measurement accuracy, the fixed interrogation pulse time is optimized for maximum excitation probability on the blue sideband after ground state cooling. The maximum delay is set such that the temperature is $\bar{n}\lesssim 1$\,ph, where this method is most sensitive when using first-order sidebands. With 200 measurement cycles, a relative accuracy of $\Delta\bar{n} /\bar{n}\approx  20$\% is reached in the range of $\bar{n}= 0.1$\,-\,$2$. For each heating-rate measurement, we typically determine the temperature at about ten different values of the $\tau$. Under normal operating conditions, heating rates of $\dot{\bar{n}}\sim1$\,ph\,s$^{-1}$ per ion are observed, which can be determined with an uncertainty of $\Delta \dot{\bar{n}} / \dot{\bar{n}} \approx 0.1$\,ph\,s$^{-1}$. The required maximum delay for such a measurement is on the order of a second, leading to a total time of around 15 minutes per heating-rate measurement.

The population in the $\ket{\mathrm{^2D}_{5/2},-5/2}$ Zeeman sublevel has a finite probability to leave the SBC cycle via decay to the long-lived $\mathrm{^2F_{7/2}}$ state. Since no fluorescence is observed in this case, the ion is detected as dark, leading to a background signal of 2.7\% after sideband cooling. As a result, the temperature measurement is limited to a minimum of $\bar{n}=0.03$ (included in the error bars). After each measurement cycle, the population is repumped from the metastable $^2\mathrm{D}_{3/2}$, the $^2\mathrm{D}_{5/2}$ and the $^2\mathrm{F}_{7/2}$ states to the ground state as shown in \fref{fig:1}(c).

\section{Heating due to dc electric field noise}\label{dc-noise}
Fluctuations of the static electric field at the ion position can change the motional state in the direction of the noise field.
In a macroscopic ion trap with an electrode-to-ion distance of $r_0$ = 0.71\,mm, the noise field originating from the electrodes  is expected to be spatially correlated at a crystal lengths $\Delta z < r_0$ \cite{Wineland:1998,King:1998} (in our case between 10 and 100\,$\mu$m). This noise leads to an energy transfer to the com mode of a chain of ions, where all ions oscillate in phase. For this process, the heating rate increases proportionally to the number of ions ($N$) and to the frequency-dependent electric field noise spectral density $S_\mathrm{E}(\omega)$, according to \cite{Home:2013,Sawyer:2014,Lechner:2016}
\begin{equation}
	\dot{\bar{n}}_{\mathrm{dc}}(\ocom) = \frac{q^2}{4 M \hbar \ocom}S_\mathrm{E}(\ocom) = N\frac{\mathrm{e}^2}{4 m_\mathrm{Yb} \hbar \ocom} S_\mathrm{E}(\ocom),
\label{eq:heatingrateDC}
\end{equation}
where $\hbar$ is the reduced Planck's constant, $\ocom$ is the secular frequency of the com mode, $q=N\times e$ is the total electric charge of the crystal, with $e$ being the elementary charge, and $M= N\times m_\mathrm{Yb}$ is the total mass of the crystal, with $m_{\mathrm{Yb}}$ being the atomic mass of ytterbium.
\subsection{Further heating mechanisms}\label{dc-mechanisms}
In modes other than the com mode, the ions oscillate out of phase relative to each other. We refer to them as out-of-phase (oop) modes and distinguish them by the number of oscillatory nodes, e.g.~the first oop mode has one node, the second oop mode has two, and so on.
Heating of the oop motional modes due to local fluctuations of electric fields on the electrodes requires gradients of these fields along the crystal and, therefore, is suppressed relative to that of the com mode by at least $\dot{\bar{n}}_\mathrm{com}/\dot{\bar{n}}_\mathrm{oop}\propto  r_0/\Delta z$ \cite{Wineland:1998,King:1998}, which in our system is on the order of $10$ or above. 

Further sources of heating can arise from the non-linearity of the Coulomb interaction via mode mixing \cite{Marquet:2003}, where phonons from two different modes can resonantly combine to an excitation in a third mode or reverse, if energy conservation is fulfilled. In a linear chain, this can be avoided by choosing trapping conditions that prevent the modes from coupling, i.e.~far from any structural phase transition (e.g.~1D$\rightarrow$2D). In contrast to the oop modes, the com mode is the only one, where the centre of mass actually moves. An external force is required to transfer energy between com and any other mode. Thus, coupling of the com mode to oop modes is suppressed\cite{Marquet:2003}. 

Motional modes can also be heated, if the ion samples anharmonicities of the trapping potential and if the resonance condition $\sum_{\alpha=1}^{3\mathrm{N}} l_\alpha\omega_\alpha = \Orf$ is fulfilled, where $l_\alpha$ is a rational number related to the order of the anharmonicity, $\omega_\alpha$ is the secular frequency of mode $\alpha$ and $N$ is the number of ions \cite{Alheit:1996,Wineland:1998}. In typical Paul trap experiments, this source of heating is avoided by choosing the Mathieu parameters $a_i, q^2_i \ll 1$ ($i = x,y,z$) and by properly choosing the secular frequencies \cite{Wineland:1998}.
\subsection{Heating of a single ion}\label{dc-single}
\begin{figure}[t]
	\begin{center}

		\includegraphics[width=\textwidth]{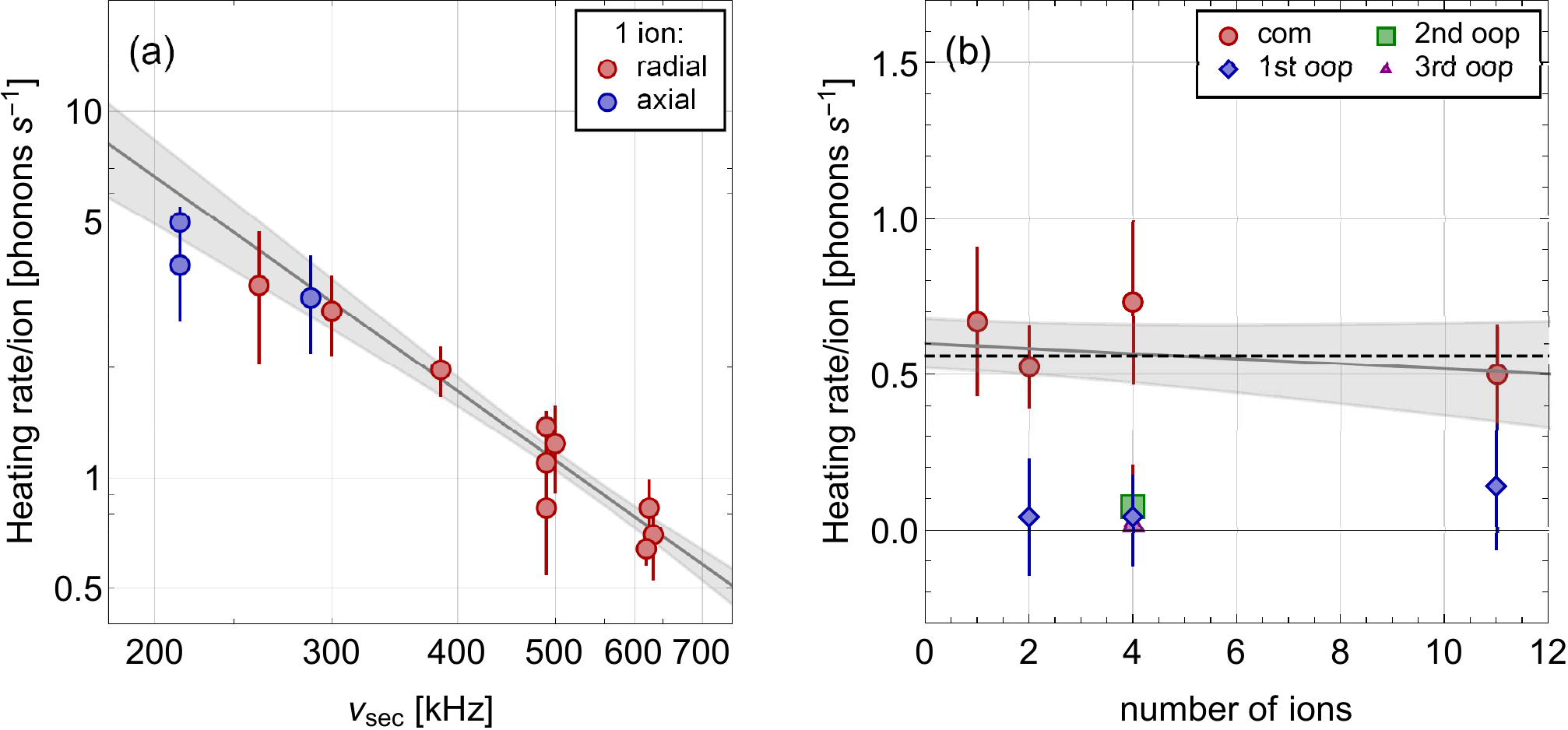}
	\end{center}
	\caption{Measured heating rates in absence of excess micromotion}{(a) Measured axial (blue circles) and radial (red circles) heating rates of a single ion as a function of the secular frequency $\nu_\mathrm{sec}=\osec/2\pi$. A power law function $\dot{\bar{n}}(\nu_\mathrm{sec})=a\nu^b_\mathrm{sec}$ is fitted to the heating rate of the radial modes (grey curve), from which $b=-1.95(25)$ is obtained. The $1\sigma$ uncertainty of the fit is indicated by the shaded grey area. The frequency dependent heating rate of the axial modes are consistent with that of the radial modes, indicating an isotropic noise distribution. \\
	(b) Heating rate per ion of the radial centre-of-mass (com) mode (red circles) and the first out-of-phase mode (blue diamonds) is shown as a function of the number of ions. A linear fit (grey curve) shows that there is no significant dependence as a function of ion number within a $1\sigma$ uncertainty of the fit (grey shaded area), therefore, we extract the weighted mean value (black dashed line) over all 4 data points of the com modes to be $\dot{\bar{n}}= 0.56(6)$\,ph\,s$^{-1}$ per ion. The secular frequencies of the com mode are in the range of $\nu_\mathrm{sec}=615$\,kHz to $\nu_\mathrm{sec}=635$\,kHz and the corresponding first oop modes are 4 to 8\,kHz lower than the com modes. For the four-ion crystal the heating rates of the other two radial out-of-phase modes at $\nu_\mathrm{sec}=619$\,kHz (green square) and at $\nu_\mathrm{sec}=607$\,kHz (purple triangle) are also measured. No significant heating of any of these modes is observed.}
	\label{fig:2}
\end{figure}

To study dc electric field noise, we measure the heating rate of the radial motional mode of a single $^{172}$Yb$^+$ ion as a function of the secular frequency $\nu_\mathrm{sec}=\osec/2\pi$ (see \fref{fig:2}, red circles). Part of the data has been published in \cite{Keller:2019b}. Data at about $\nu_\mathrm{sec}=600$\,kHz to $\nu_\mathrm{sec}=630$\,kHz was recorded repeatedly and shows consistent results over the course of four years. A power law function $\dot{\bar{n}}(\nu_\mathrm{sec})=a\nu_\mathrm{sec}^b$ is fitted (grey line) to the data of the radial mode (red circles) as shown in \fref{fig:2}(a), yielding $b=-1.95(25)$, in agreement with what is found in other traps, see e.g.~\cite{Boldin:2018}. By fixing the exponent to $b=-2$, we obtain a heating rate of $\dot{\bar{n}}\left(\nu_\mathrm{sec}\right)=2.88(17)\times10^{11}\mathrm{s^{-1}Hz}^2/\nu_\mathrm{sec}^2$, which corresponds to an electric field noise spectral density of $S_\mathrm{E}\left(\nu\right)=8.49(8)\times10^{-9} \mathrm{(V/m)}^2/\nu$, according to \eref{eq:heatingrateDC}. 
In addition, the heating rate of the axial motional mode is measured for secular frequencies between $\nu_ \mathrm{sec}= 200$\,kHz and $\nu_ \mathrm{sec}= 300$\,kHz and is found to be consistent with this electric field noise spectral density, as shown in \fref{fig:2}(a) (blue circles). Currently, the operating voltages of $\leq 12$\,V provide a maximal axial secular frequency of $\nu_\mathrm{sec} \approx 300$\,kHz. The heating rate of the axial mode as a function of secular frequency is in agreement with that of the radial mode, indicating that the DC noise field is isotropic.

\subsection{Heating of linear crystals} \label{dc-two}
During sideband thermometry of larger ion chains, the spectroscopy laser at 411\,nm addresses all the ions globally, producing entangled states when driving motional sidebands, while detection of the electronic state is done on one of the outer ions in the chain. If the average motional state occupation is low, i.e.~the number of phonons is smaller than the number of ions, the probability of a red sideband excitation of one of the ions depends on the state of the other ions. We numerically calculate a correction factor for the extracted value of $\bar{n}$ from the detected sideband ratio, which takes the correlation between the ions into account \cite{Home:2013,Home:2000,King:1998}. A more detailed description is reported in \ref{appendixA}.

The heating rates of radial modes of linear crystals of two, four and eleven ions are measured at a fixed radial confinement, see \fref{fig:2}(b). The axial potential is adapted to the crystal length for several reasons. Firstly, the critical axial secular frequency for the 1D to 2D transition (linear-to-zigzag \cite{Birkl:1992,Schiffer:1993,Dubin:1993}) reduces with increasing number of ions. Therefore, if the radial secular frequency is kept constant, the axial secular frequency should be kept low enough to prevent cross-coupling of modes. Secondly, the frequency splitting of the radial modes is proportional to the axial secular frequency and a reduction of the axial secular frequency, therefore, facilitates simultaneous sideband cooling of several modes. Note, that this is limited by the ability to sufficiently resolve individual motional modes during sideband thermometry. Using the RSC method discussed in \sref{experiment}, all radial modes of the two-ion and four-ion crystal were cooled to $\bar{n} \leq 0.1$. For the 11-ion crystal, the detuning of the spectroscopy laser at 411\,nm is set such that the four modes with the highest secular frequency are cooled efficiently. To exclude heating from non-linear coupling between sideband cooled modes and Doppler cooled modes in the investigated crystals, we perform molecular dynamics simulations as described in \ref{appendixB}. With this, we numerically verify, that under the experimental conditions there is no significant energy flow to the cooled modes from any other mode. The calculated energy fluctuations in the sideband-cooled modes due to non-linear coupling are below $10^{-4}$\,ph.

The heating rate of the com mode is measured to be $\dot{\bar{n}} = 0.56(6)$\,ph\,s$^{-1}$ per ion, as can be seen from the black dashed line. The data is fitted to a linear function (grey line), verifying that the heating rate \textit{per ion} is constant as function of ion number within the uncertainty (shaded grey area). The heating rate of the first oop mode is consistent with zero within the uncertainty of $\sigma_{\dot{\bar{n}}}=0.2$\,ph\,s$^{-1}$ per ion for all crystal sizes, ranging between $10\,-100\,\mathrm{\mu m}$. Heating rates of the third and fourth oop mode in the four-ion crystal are also measured and no significant heating is observed at the level of our resolution.

\FloatBarrier
\section{RF electric field noise coupling to secular motion}\label{rf-noise}
When the ion is exposed to EMM, external and technical noise at Fourier components $\Orf \pm \osec$ couple to the ion motion, leading to additional heating at secular frequency $\osec$ \cite{Wineland:1998,Blakestad:2009}. The ponderomotive trapping potential \cite{Dehmelt:1967}

\begin{equation}
\Phi_\mathrm{P} = \frac{q^2}{2m\Orf^2} \times \left< \vec{E}^2_\mathrm{rf}(x,y,z,t)\right>, \label{eq:phi}
\end{equation}
where $m$ is the ion mass, $q$ is the ion charge, $\Orf$ is the rf drive frequency, $\vec{E}_\mathrm{rf}(x,y,z,t)$ the applied electric field, which is time averaged over a full rf period of $T=2\pi/\Orf$. Assume a trapping electric field with a small noise contribution $\xi$ at $\Orf \pm \osec$, 
\begin{equation}
\vec{E}_\mathrm{rf}(\vec{r},t) = \vec{E}_0(\vec{r})\left\{\cos(\Orf t) + \xi\cos[(\Orf \, \pm \, \osec)t]\right\}.
\end{equation}
The time average of \eref{eq:phi} reproduces the confining pseudopotential and, additionally, yields a $\cos(\osec t)$~-~dependent beating term. The gradient of the beating term provides a periodic noise force at frequency $\osec$ acting on the ion motion
\begin{equation}
	\vec{F}= -\frac{q^2}{2m\Orf^2}\vec{\nabla}E^2_\mathrm{0}(\vec{r})\xi \cos(\osec t).
\end{equation}
The noise force spectral density $S_\mathrm{F}$ is related to the voltage noise spectral density via  $S_\mathrm{F}/F^2 = S_\mathrm{V}/V_0^2 =S_\mathrm{E}/E_0^2$, providing an extension to \eref{eq:heatingrateDC} for the rf-induced heating rate of mode $\alpha$ with secular frequency $\omega_\alpha$. The total heating rate is then given by $\dot{\bar{n}}_{\mathrm{total}}(\omega_\alpha) = \dot{\bar{n}}_{\mathrm{dc}}(\omega_\alpha)+\dot{\bar{n}}_{\mathrm{rf}}(\omega_\alpha)$, where
\begin{equation}
\dot{\bar{n}}_{\mathrm{rf}}(\omega_\alpha) = \frac{1}{4m\hbar\omega_\alpha} S_\mathrm{F}(\omega_\alpha) = \frac{1}{4m\hbar\omega_\alpha} F^2_\alpha \frac{S_\mathrm{V}(\Orf\pm\omega_\alpha)}{V_0^2}.
\end{equation}
Since the ion senses always both contributions, we set here $S_\mathrm{V}(\Orf\pm\omega_\alpha) = S_\mathrm{V}(\Orf+\omega_\alpha) + S_\mathrm{V}(\Orf-\omega_\alpha)$.\\
Decomposing the ion oscillations in the trapping potential into normal modes of motion \cite{James:1998}, the modes are described by sets of eigenfrequencies $\omega_{\alpha}$ and normalized eigenvectors $\vec{\beta}_\alpha$. In this formalism, the relative motional amplitude of ion $j$ in mode $\alpha$ is given by the $j$-th component of the eigenvector $\vec{\beta}_\alpha$.
In general, a mode $\alpha$ with a normalized mode vector $\vec{\beta}_\alpha$ will experience a noise force $F_\alpha = \vec{F}(\vec{r})\cdot \vec{\beta}_{\alpha}$ according to its projection on the force vector. For a single ion with its mode direction parallel to the gradient of the pseudopotential, the relation for the heating rate is given by \cite{Blakestad:2009}
\begin{equation}\label{eq:heatingrateRF}
	\dot{\bar{n}}_{\mathrm{rf}}(\omega_\alpha) =  \frac{e_\mathrm{0}^4}{16m^3\Orf^4\hbar \omega_\alpha} \left[\vec{\nabla} E^2_\mathrm{0}(\vec{r})\right]^2 \frac{S_\mathrm{V}(\Orf\pm\omega_\alpha)}{V_\mathrm{0}^2}.  
\end{equation}
Thus, measuring the heating rate as function of $\vec{\nabla} E^2_\mathrm{rf}(\vec{r}) \propto r$ provides access to a part of the voltage noise spectrum at the ion position.
\FloatBarrier
\subsection{Heating of a single ion}\label{rf-single}
\begin{figure}[tb]
	\begin{center}

	\includegraphics[width=\textwidth]{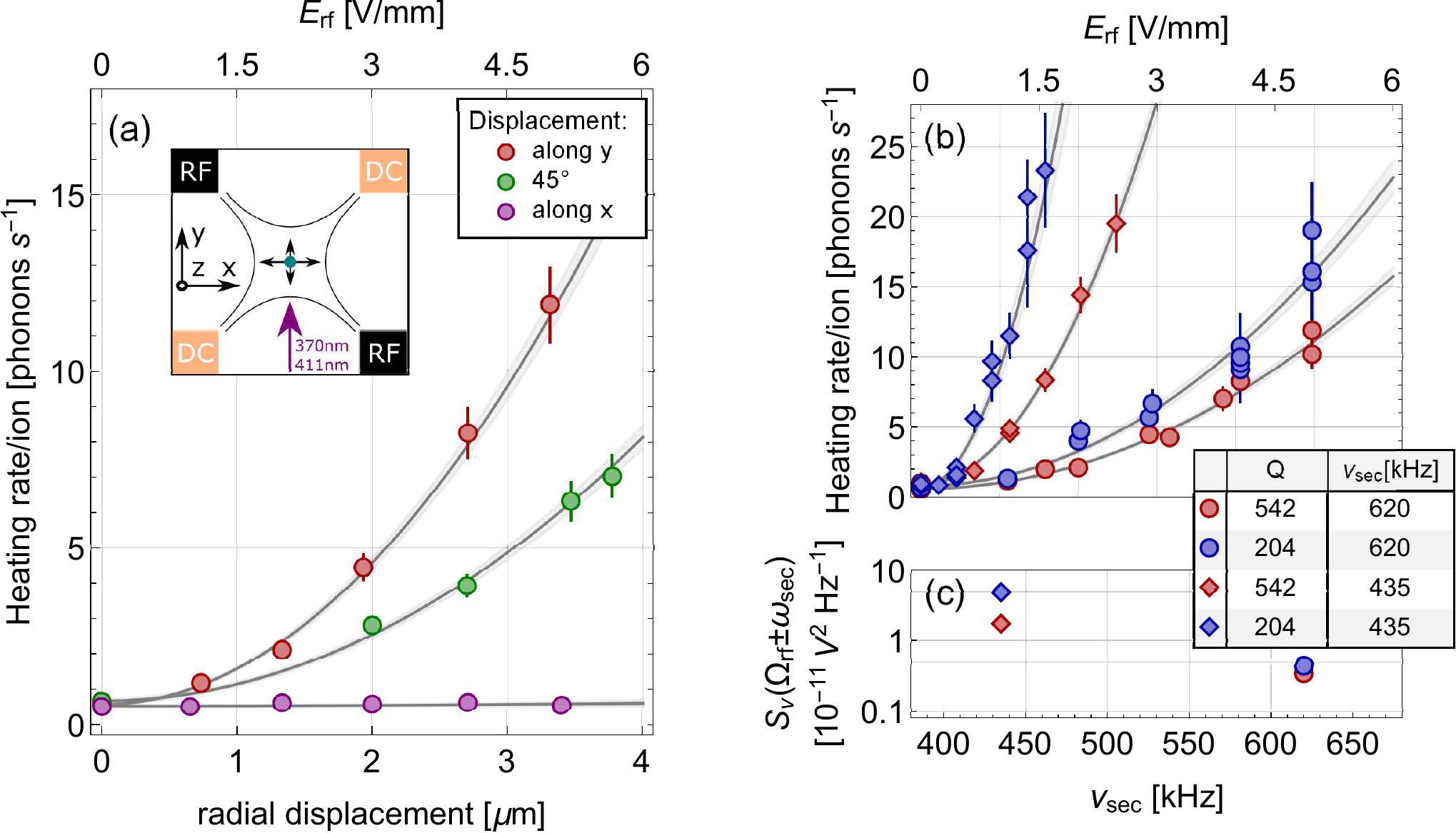}
	\end{center}
	\caption{Heating under influence of excess micromotion (EMM)}{
	(a) Heating rate of a single-ion radial mode when exposed to radial EMM. The inset shows the orientation of the principle axes in the trap and the direction of the spectroscopy laser. The ion is displaced along $y$ (red circles), along $x$ (purple circles) and diagonally in the plane (green circles). The data is fitted (grey curves) to equation \eref{eq:heatingrateRF} and the 1$\sigma$-uncertainty is shown by the grey shade.\\
	(b) Heating rate of the radial mode of a single ion as a function of displacement for two different values of the quality factor ($Q$) of the resonant circuit. Data is recorded at $Q=542$ (red data points) and $Q=204$ (blue data points) and at a secular frequency $\nu_\mathrm{sec}= 435$\,kHz (diamonds) and $\nu_\mathrm{sec}= 620$\,kHz (circles).\\
	(c) Extracted voltage noise spectral density $S_\mathrm{V}$ from the curves shown in (b). A significant increase of $S_\mathrm{V}$ is observed at a reduced $Q$ and $\nu_\mathrm{sec}$. }
	\label{fig:3}
\end{figure}
To measure the heating rate of an ion under the influence of rf noise, the experimental sequence is modified to radially displace the ion by $r$, inducing excess micromotion. For this, the compensation voltages $U_\mathrm{tc}$ and $U_\mathrm{ec}$ are changed, such that the ion is displaced radially by a few $\mu$m and stored at a specific $E_\mathrm{rf}$ ($\propto r$). After the delay $\tau$, the ion is returned to the nodal line and the temperature of the mode is determined in the same manner as described in \sref{experiment}. 
\begin{table}[btp]
	\centering
	\caption{Relations between applied voltages $\Delta U_\mathrm{tc}, \Delta U_\mathrm{ec}$, electric fields $\mathrm{E_{rf}}$ and displacement $\Delta x$, $\Delta y$ for a $^{172}$Yb$^+$ ion at radial secular frequency of $\nu_{sec}$ = 620\,kHz. Additionally, the corresponding displacing static electric field $\mathrm{E_{dc}}$ is given in columns 7 and 8.}
	\begin{tabular}[h]{cc|cccccc} \label{table-emmslopes}
		$\Delta x$&$\Delta y$&$\Delta U_\mathrm{tc}$&$\Delta U_\mathrm{ec}$&$\mathrm{E_{rf, x}}$&$\mathrm{E_{rf, y}}$&$\mathrm{E_{dc, x}}$& $\mathrm{E_{dc, y}}$ \\ [1pt]
		[$\mu$m]&[$\mu$m]&[mV]&[mV]&[V/mm]&[V/mm]&[V/m]&[V/m]  \\ [1pt]
		 \hline \hline
					1 &  - & 33 &-165& 1.505 & 0 & 0 & 27.0  \\
					- &  1 & -2.9 & 278 & 0 & 1.505 & 27.0 & 0 \\
					0.707 &  0.707 & 21.5 & 79.7 & 1.063 & 1.063 & 19.1 & 19.1 \\
	\end{tabular}
\end{table}

$E_\mathrm{rf}$ was measured up to $E_{\mathrm{rf}}\approx 1$\,V mm$^{-1}$ once as a function of the applied compensation voltages to obtain the four relations $\partial E_\mathrm{rf, x} / \partial U_\mathrm{tc}$, $\partial E_\mathrm{rf, y} / \partial U_\mathrm{tc}$, $\partial E_\mathrm{rf, x} / \partial U_\mathrm{ec}$ and $\partial E_\mathrm{rf, y} / \partial U_\mathrm{ec}$ with uncertainties below 2\%, using photon correlation method \cite{Berkeland:1998,Keller:2015a}.
From day to day, the stray electric fields are compensated within $\sigma_\mathrm{E_{rf}}=  0.05$\,V\,mm$^{-1}$ and the ion is displaced in the $xy$ plane using a combination of voltages $U_\mathrm{tc}$ and $U_\mathrm{ec}$, where we extrapolate for $E_{\mathrm{rf}} \gtrsim 1$\,V mm$^{-1}$. \Tref{table-emmslopes} shows exemplary values for $\nu_\mathrm{sec}=620$\,kHz.

The low-pass filters connected to the DC electrodes suppress frequencies higher than 113\,Hz \cite{Keller:2019b}, resulting in a characteristic rise and fall time of 9\,ms for externally applied voltage changes. This ensures adiabatic transport of the ions as voltages are adjusted on much longer time scales than an oscillation period of the ion. We verify that the displacement itself does not heat the investigated mode by repeatedly shifting the ion outwards and back without any additional  delay ($\tau=0$) and measuring the temperature. Accounting for additional small delays in experiment control, the total duration of the displacement sequence (back and forth) is about 28\,ms.  Therefore, we extend $\tau$ by a fixed additional delay time $\tau_\mathrm{tech}=40$\,ms, so that the experiment remains synchronous with the 50\,Hz mains line cycle. As discussed in \sref{experiment}, the highest measurement sensitivity is reached at values $\bar{n}\lesssim1$. Therefore, the total $\tau_\mathrm{total} = \tau + \tau_\mathrm{tech}$ is adapted at each displacement to account for the expected heating rate.

\Fref{fig:3}(a) shows the measured heating rate of the radial mode along the $y$ direction at $\osec=2\pi\times620$\,kHz as a function of displacement in three directions. First, the ion is displaced along $y$, so that the measured mode is parallel to $\vec{\nabla} E^2_\mathrm{rf}$ (red points). A quadratic increase of the heating rate is observed as a function of the displacement. A fit to the relation $\dot{\bar{n}}(r)=Ar^2$ yields $A = 1.03(1)$\,ph\,s$^{-1}$\,$\mu$m$^{-2}$, corresponding to a voltage noise spectral density of $S_\mathrm{V}(\Orf \pm \osec) = 0.37(1) \times 10^{-11}$\,V$^2$Hz$^{-1}$ according to \eref{eq:heatingrateRF}. Next, the ion is displaced diagonally in the $xy$ plane (green points) at an angle of $\phi =45(3)^{\circ}$, where less heating is observed as a function of displacement. The fit yields a quadratic coefficient $A=0.45(2)$\,ph\,s$^{-1}$\,$\mu$m$^{-2}$. From the projection of the $y$-mode on the radial gradient $\left[\vec{F}_\mathrm{y}(\vec{r})\cdot \vec{\beta}_\mathrm{y}\right]^2 \propto \cos^2(\phi)$, a factor of $0.50(5)$, i.e.~a coefficient $A=0.52(5)$\,ph\,s$^{-1}$\,$\mu$m$^{-2}$, is expected for a diagonal displacement. Finally, when shifting the ion along $x$, where the gradient is perpendicular to the mode vector, no additional heating of the radial $y$-mode due to rf noise is observed (purple points). It is worth noting, that the total energy transfer from the noise field to the ion is unchanged, but is differently distributed over the two radial modes.

\subsection{Impact of the quality factor of the rf circuit}\label{rf-qualityfactor}
As the ion motion couples to noise at $ \Orf \pm \osec$ under the influence of EMM, this type of heating is expected to be strongly dependent on  both, the quality factor ($Q$) of the resonant rf circuit that drives the confining rf field and on the secular frequency of the interrogated mode.

In order to study this effect, the heating rate of a single ion is measured under the influence of EMM in two different configurations. At typical operating conditions, where the trap drive frequency is $\Orfind{1}=2\pi \times24.4$\,MHz, the Lorentzian-shaped rf resonance has a full width at half maximum (FWHM) of $45$\,kHz, corresponding to a loaded quality factor of $Q_1=542$ (unloaded $Q=1055$). The measurement is repeated with a reduced quality factor of $Q_2=204$. Besides the increased FWHM = $116$\,kHz, the resonance frequency shifts to $\Orfind{2}=2\pi \times23.7$\,MHz. To retain the same secular frequency of the ion to the initial configuration, the power of the rf source is increased.
For each value of $Q$, data is recorded at two different secular frequencies of $\osecind{1} =2\pi\times 620$\,kHz and $\osecind{2} = 2\pi\times435$\,kHz, as is shown in \fref{fig:3}(b). As expected, no influence of $Q$ is observed on the heating rate at compensated micromotion, i.e.~within $\sigma_{\mathrm{E_{rf}}}=0.05$\,V\,mm$^{-1}$. However, when the ion motion couples to rf noise through EMM, a significant increase of the heating rate is observed, as can be seen from the slope of the curves in \fref{fig:3}(b). This effect becomes larger at lower secular frequency, because then the noise at $\Orf \pm \osec$ is amplified more by the resonant circuit. From the fits, the voltage noise spectral density $S_\mathrm{V}(\Orf \pm \osec)$ is determined for $Q_{1/2}$ and $\osecind{1/2}$, see \fref{fig:3}(c), and yields, e.g.,~$S_\mathrm{V}(\Orfind{1} \pm \osecind{1})=0.340(12)$\,V$^2$\,Hz$^{-1}$. At $Q_2 = 205$, the voltage noise is amplified by a factor $1.28(9)$ for a secular frequency of $\osecind{1}=2\pi\times620$\,kHz, while a factor of $2.80(15)$ is observed for $\osecind{2}=2\pi\times435$\,kHz. 
Using the Lorentzian relation for transmission of the helical resonator \eref{eq:helicalTransmission}, a power spectral density of the noise is determined to be $S_\mathrm{P} < 4\times10^{-14}$\,W\,Hz$^{-1}$, comparable with values obtained in other systems, see e.g.~\cite{Sedlacek:2018}. The extracted values of $S_\mathrm{P}$ at the four different frequencies $\Orfind{1,2} \pm \osecind{1,2}$ are significantly different, showing that the noise is not white in this range.
\FloatBarrier
\subsection{Heating of a two-ion crystal}\label{rf-two}
\begin{figure}[t]
	\begin{center}
		\includegraphics[width=\textwidth]{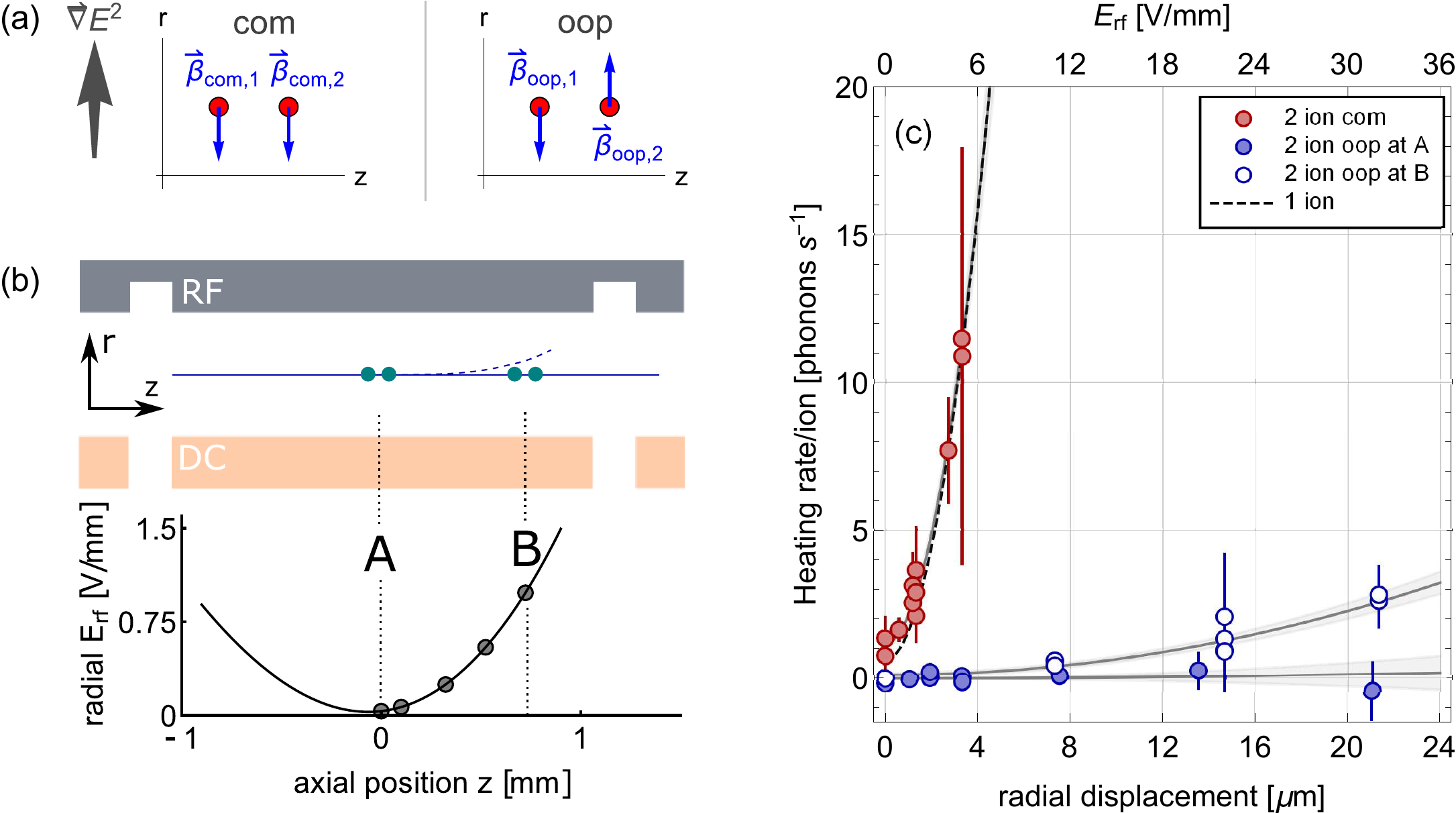}

	\end{center}
	\caption{Heating rate under EMM of a spatially extended  crystal}{
	(a) Radially displaced two-ion crystal and the vectors (blue arrows) of the com mode (left) and oop mode (right), together with the pseudopotential gradient ($\vec{\nabla} E^2$) (grey arrows).\\
	(b) Schematic representation of the investigated trap segment (top) and measured excess micromotion (error bars are smaller than symbols) along with a spline to guide the eye (bottom). The positions in the trap segment, where heating-rate measurements are performed are indicated by A (centre) and B (edge). In the trap schematic, the solid line indicates the rf nodal line and ion positions at compensated EMM. The dashed blue line exaggerates the deviation of the ion position from the nodal line, when they are shifted to position B (see text).\\
	(c) Heating rate of the radial com and oop mode when exposed to EMM in the centre of the trap segment (A, full blue circles) and at the edge of the trap segment (B, empty blue circles). In the latter position, heating of the oop mode is observed due to a significant axial gradient of the radial electric field. The single ion data set reproduces consistently the one from \fref{fig:3}, which was recorded about five months earlier.}
	\label{fig:4}
\end{figure}

In order to investigate the impact of the rf field gradient on extended ion Coulomb crystals, we first consider the most simple example of a two-ion crystal. \Fref{fig:4}(a) shows an axial crystal of two ions and the mode vectors (blue arrows) of the radial com mode and the radial oop mode, together with the pseudopotential gradient ($\vec{\nabla} E_\mathrm{rf}^2$), which is identical for both ions (grey arrows). The energy transfer to the com mode is expected to be twice that of a single ion as the force on the mode is doubled. As a result, the heating rate per ion should be equal. For the oop mode though, the mode vectors $\vec{\beta}_{\alpha,j}$ are oriented in opposite direction ($\vec{\beta}_{\mathrm{oop},1} = -\vec{\beta}_{\mathrm{oop},2}$) leading to a cancellation of the noise force, i.e.~the net force on the mode is zero, and thus no heating of the oop mode should occur. In fact, this is the case for all radial oop modes in larger strings of ions, which are radially displaced, as pairwise cancellation of the noise force occurs. Or more fundamentally, for each oop mode ($\alpha \neq \mathrm{com}$), the sum of the mode vectors components $\vec{\beta}_{\alpha,j}$ in positive direction is equal to the sum of those in the negative direction, since the centre of mass is at rest in those modes. As the ions are placed in an identical gradient $ \vec{\nabla} E_\mathrm{rf}^2(\vec{r}_j) = \vec{\nabla} E_\mathrm{rf}^2(r)$, the force contributions must cancel
\begin{equation}\label{eq:fothermodes}
F_{\alpha \neq \mathrm{com}} \propto \sum_{j=1}^{N}\vec{\nabla} E_\mathrm{rf}^2(\vec{r}_j)\cdot \vec{\beta}_{\alpha,j} = \vec{\nabla} E_\mathrm{rf}^2(r)\cdot\sum_{j=1}^{N} \vec{\beta}_{\alpha,j} = 0
\end{equation}
 As a result, only the radial com mode of an ion chain should be heated by rf noise.
 
To experimentally study the influence of rf noise on a larger crystal, the heating rate of a linear chain of two ions at a separation of 17\,$\mu$m is measured as a function of radial displacement in the same way as described in \sref{rf-single}. The results are shown in \fref{fig:4}(c), where the red points show the heating rate of the com mode at around $\osec=2\pi\times620$\,kHz, and the filled blue points show the heating rate of the oop mode at around $\osec=2\pi\times615$\,kHz. The data is fitted to $\dot{\bar{n}} = A r^2$, yielding  $A = 0.89(4)$\,ph\,s$^{-1}$ $\mu$m$^{-2}$ per ion for the com mode, which is in agreement with that of a single ion (the single-ion fit is given by the black dashed line) and confirms the expected scaling of the heating rate with ion number. The grey shaded area indicates the $1\sigma$ uncertainty of the fit. As expected from \eref{eq:fothermodes}, no significant heating of the oop mode is observed (blue full symbols in \fref{fig:4}(c)) up to a displacement $\Delta r = $ 21\,$\mu$m.

In order to induce heating by rf noise of the oop mode, we apply voltages $\mathrm{U_{tc}, U_{ec}}$ in the neighbouring segment. In this manner, the radial component of the rf electric field as seen by the ions near the segment edge strongly depends on the $z$ coordinate, as shown in \fref{fig:4}(b), i.e.~the ions will deviate from the nodal line while shifting them along $z$. Next, the ions were moved from the segment centre (position A in \fref{fig:4}(b)) by 720\,$\mu$m along the axial trapping direction towards the edge of the segment (position B in \fref{fig:4}(b)), keeping the ion separation constant. The difference in the pseudopotential gradient at the ion positions results in a different coupling of noise to the individual ions, preventing the noise force from fully cancelling. This differential noise coupling becomes larger as we displace the ions radially and leads to significant heating of the oop mode at high values of EMM (open blue circles in \fref{fig:4}(c)). With respect to the heating rate of the com mode, $16.7$ times less heating is observed in the case of the oop mode.

\FloatBarrier

\section{RF-noise induced heating of radially extended crystals}\label{extended-crystals}
Based on the observed rf-noise induced heating of secular motion in few-ion systems under the presence of rf field gradients, we extend the discussion now by theoretical calculations of more complex  2D and 3D crystals. Here, we focus on the heating effect due to rf noise and neglect the known heating effect due to non-linearity of the Coulomb potential, which becomes relevant at higher temperatures \cite{Ryjkov:2005}. The ground state configuration of an N-ion crystal is calculated at temperature $T=0$\,K under high damping using molecular dynamic simulations. The chosen trapping potential is approximated to the second-order and sets of $3N$ eigenfrequencies $\omega_\alpha$ and mode vectors $\vec{\beta}_{\alpha,j}$ are obtained from solving for the eigensystem.

In the case of 2D or 3D crystals, the mode structure becomes complicated and the mode vectors of individual ions are typically no longer oriented along a single principal axis of the trap. Therefore, the total noise force ($F_\alpha$) on mode $\alpha$ is obtained by summing over the forces acting on each individual ion $j$, according to
\begin{equation}
	F_\alpha = \sum_{j=1}^{N} \vec{F}_{j}(\vec{r}_j)\cdot\vec{\beta}_{\alpha,j},
\end{equation}
which is proportional to the projection of its normalized mode vector $\vec{\beta}_{\alpha,j}$ on the gradient of the electric field.
This modifies equation \eref{eq:heatingrateRF} to
\begin{equation}\label{eq:heatingrateRFmulti}
	\dot{\bar{n}}_{\mathrm{rf}}(\omega_\alpha) =  \frac{e_0^4}{16m^3\Orf^4\hbar \omega_\alpha} \left[\sum_{j=1}^{N} \vec{\nabla} E^2_\mathrm{rf}(\vec{r}_j)\cdot\vec{\beta}_{\alpha,j}\right]^2 \frac{S_\mathrm{V}(\Orf\pm\omega_\alpha)}{V_\mathrm{0}^2},
\end{equation}
 from which the heating rate on a mode $\alpha$ can be obtained. Since $\vec{\nabla} E^2_\mathrm{rf}(\vec{r}_j)$ is proportional to the absolute ion distance $r_j$ from the nodal line, the square of the term is dominated by contributions from the outermost ions at radial distance around $r_\mathrm{max}$ from the nodal line. Therefore, we obtain an approximate scaling of the heating rate $\dot{\bar{n}}_{\mathrm{rf},\alpha}\propto  \omega_\alpha^{-1} \times r_\mathrm{max}^2$.
 
 For simplicity, we assume a constant power spectral density of $S_\mathrm{P} = 1.5\times10^{-14}$\,W/Hz over the range of frequencies $\Orf \pm \omega_\alpha$ at the input of the resonant rf circuit. This value is in the range of the experimentally obtained values, see \sref{rf-single}. Using a Lorentzian transfer function, the spectral voltage noise $S_\mathrm{V}(\Orf \pm \omega_\alpha)$ on the rf electrode is calculated \cite{Sedlacek:2018}
\begin{equation}\label{eq:helicalTransmission}
	S_\mathrm{V}(\Orf\pm\omega_\alpha) = \frac{Q L \Orf}{1 + 4 Q^2 (\frac{\omega_\alpha}{\Orf})^2} S_\mathrm{P}(\Orf\pm\omega_\alpha).
\end{equation}
The transmission is dependent on the quality factor $Q=542$ and the inductance $L=2.5$\,$\mu$H of the resonant circuit and scales approximately as $\omega_\alpha^{-2}$. Using \eref{eq:heatingrateRFmulti}, we obtain $\dot{\bar{n}}_{\mathrm{rf},\alpha}\propto \omega_\alpha^{-3} \times r^2_\mathrm{max}$.

\begin{figure}[b]
	\begin{center}

		\includegraphics[clip,width=\textwidth]{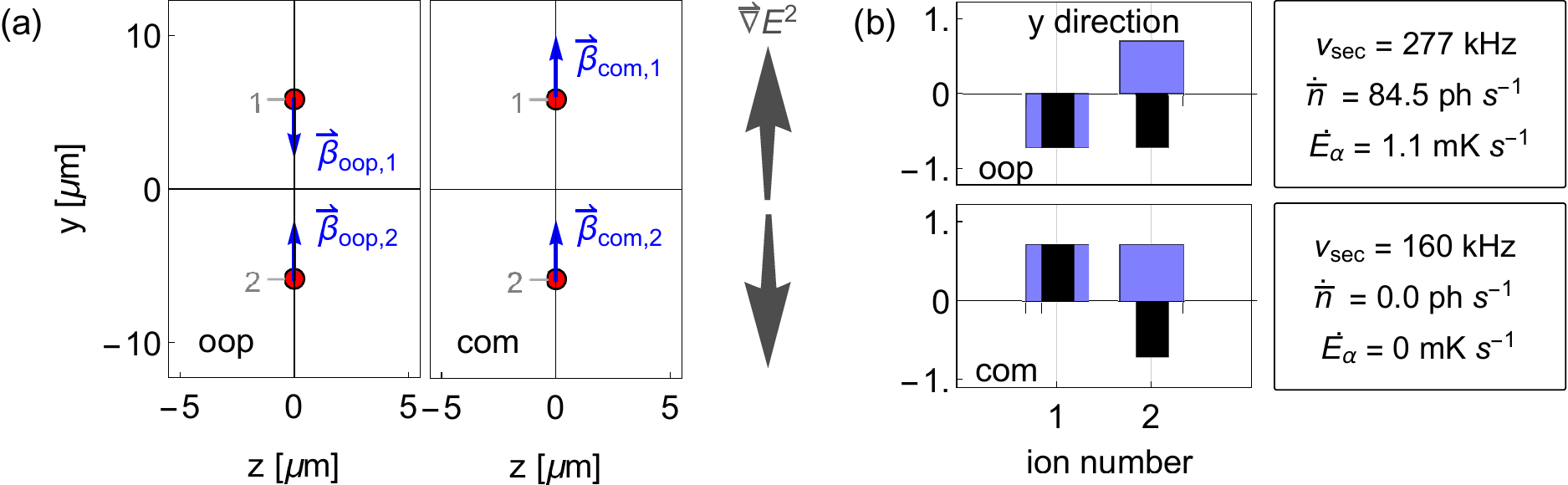}
	\end{center}
	\caption{Heating of a radially extended crystal}{
		(a) A radially extended two-ion crystal for $(\omega_\mathrm{x},\omega_\mathrm{y},\omega_\mathrm{z}) = 2\pi\times(225,\,160,\,223)$\,kHz. The mode vectors of the out-of-phase (oop) mode and for the centre-of-mass (com) mode are depicted with blue arrows. The direction of the gradient of the pseudopotential ($\vec{g}=\vec{\nabla}E_\mathrm{rf}^2$) is shown with thick grey arrows.\\
		(b) Normalized mode vector amplitudes $\beta^y$ in the $y$ direction (blue shaded bars) and individual force contributions $\vec{g}_j\cdot  \vec{\beta}_{j}$ on the mode (black bars) for both the oop mode and the com mode. The mode frequency $\nu_\mathrm{sec}=\osec/2\pi$, the calculated heating rate $\dot{\bar{n}}$ in ph\,s$^{-1}$ and $\dot{E}$ in mK\,s$^{-1}$ of the different modes are given in the textboxes on the right.}
	\label{fig:5}
\end{figure}
\subsection{Radially oriented linear crystals}\label{two-ion-radial}
To understand the behaviour of radially extended crystals, we first consider a simple two-ion crystal oriented along a radial direction ($y$). Please note, that from here on, we make the choice of the weaker confinement to be along the $y$ direction.  We assume an ideal linear ion trap, where no axial rf fields or gradients are present, and only focus on the radial components of the modes. \Fref{fig:5}(a) depicts the calculated ion positions for a confinement of $(\omega_\mathrm{x},\omega_\mathrm{y},\omega_\mathrm{z}) = 2\pi\times(225,\,160,\,223)$\,kHz. The blue arrows indicate the direction of mode vectors of the radial oop mode at $\osec=2\pi\times277$\,kHz and the radial com mode at $\osec=2\pi\times160$\,kHz. The two ions are symmetrically placed around the nodal line, at a distance of 5.8\,$\mu$m. The pseudopotential gradient ($\vec{g}_j = \vec{\nabla} E_\mathrm{rf}^2(\vec{r}_j)$), coupling the noise to each ion, points in opposite directions with respect to the nodal line ($\vec{g}_1 = -\vec{g}_2$), as depicted by grey arrows. The mode vectors of the oop mode also point in opposite directions and are of equal amplitude ($\vec{\beta}_{\mathrm{oop},1} = -\vec{\beta}_{\mathrm{oop},2}$). Therefore, the individual noise force contributions $\vec{g}_1\cdot  \vec{\beta}_{\mathrm{oop},1} + \vec{g}_2 \cdot \vec{\beta}_{\mathrm{oop},2}$ add constructively. This is visualized in \fref{fig:5}(b) with black bars showing the normalized individual force contributions and blue bars showing the normalized mode amplitudes for each mode along the $y$ direction. In this case, the oop mode is heated by 84.5\,ph\,s$^{-1}$, corresponding to $\dot{E} = \dot{\bar{n}}\hbar\omega/k_B = 1.1$\,mK\,s$^{-1}$. 

In contrast, the mode vectors of the com mode point in the same direction  ($\vec{\beta}_{\mathrm{com},1} = \vec{\beta}_{\mathrm{com},2}$). Therefore, the force contributions of the individual ions have opposite sign, as seen from the black bars in the lower panel of \fref{fig:5}(b), and the net force on this mode is zero ($\vec{g}_1\cdot  \vec{\beta}_{\mathrm{com},1} + \vec{g}_2 \cdot \vec{\beta}_{\mathrm{com},2} = 0$). As a result, the com mode of the extended two-ion crystal is protected from rf-noise induced heating. This perfect cancellation only holds if the ions are symmetrically placed around the nodal line. In the case of uncompensated stray electric fields, the net force on the com mode is finite and it is also heated.
\begin{figure}[b]
	\begin{center}

		\includegraphics[clip,width=\textwidth]{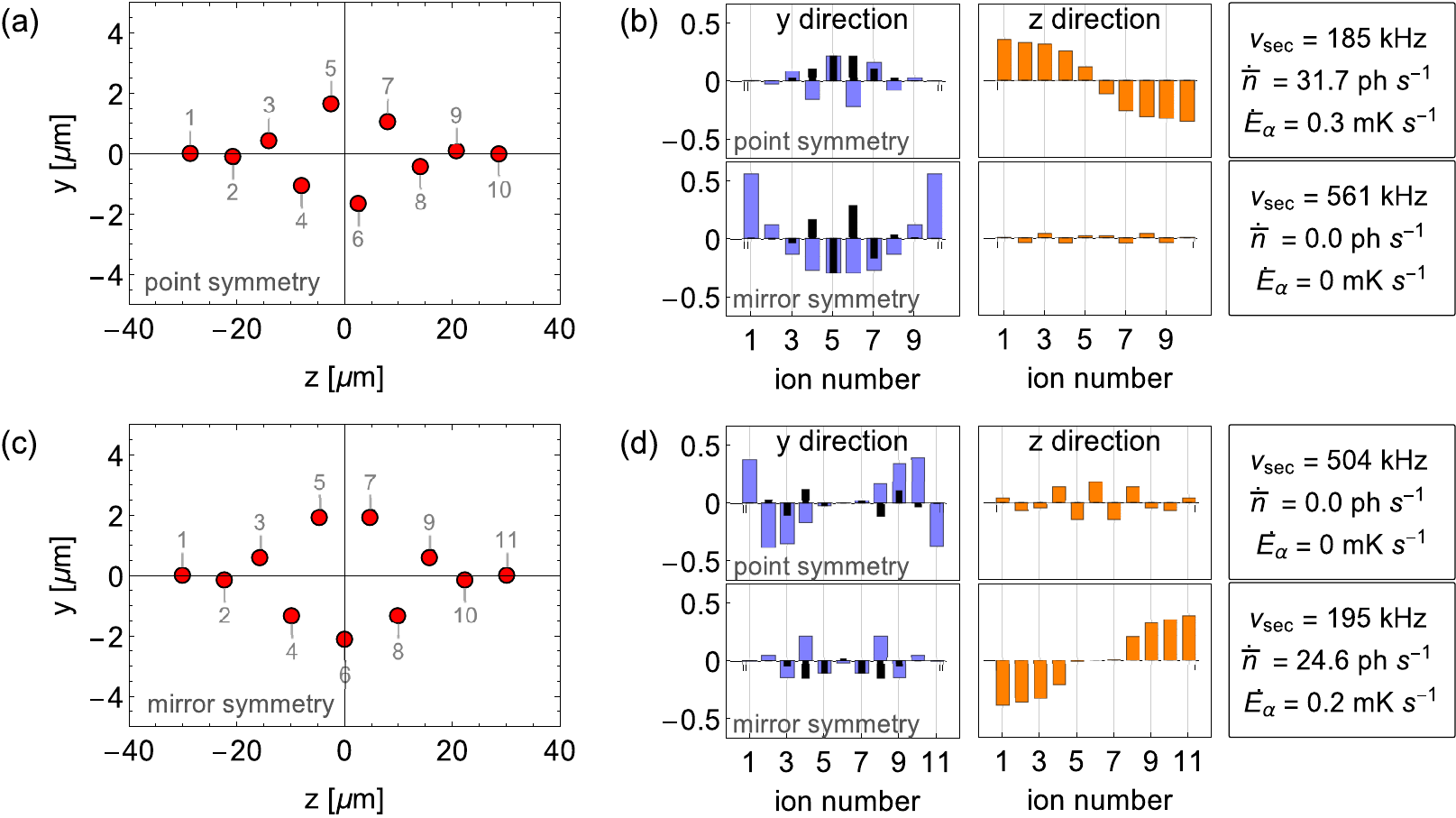}
	\end{center}
	\caption{Heating of 2D extended crystals}{
		Ion positions of a (a) 10-ion and (c) an 11-ion 2D crystal for $(\omega_\mathrm{x},\omega_\mathrm{y},\omega_\mathrm{z}) = 2\pi\times(601, 600, 140)$\,kHz. The respective normalized mode amplitude ($y$-component blue, $z$-component orange) and force contribution (black bars) in the $y$ and $z$ direction are shown in (b) and (d) together with the mode frequency and the heating rate in terms of both phonon number and energy. The 10-ion crystal is \textit{point-symmetric} with respect to the crystal center, while the 11-ion crystal is \textit{mirror-symmetric} with respect to the $y$ axis ($z=0$). Heating of radial modes occur if the symmetry of the mode is of the same kind as the symmetry of the crystal (see text). For each crystal, the upper panel of (b) and (d) shows a mode with radial point symmetry while the lower panel shows a mode with radial mirror symmetry, respectively. For both crystals the most heated mode is shown as well as an example of a protected mode. 
		
	}
	\label{fig:6}
\end{figure}
\subsection{2D zig-zag crystals}\label{2Dcrystal}
The calculated ion positions of a 10- and an 11-ion zig-zag crystal for trap frequencies $(\omega_\mathrm{x},\omega_\mathrm{y},\omega_\mathrm{z}) = 2\pi\times(601, 600, 140)$\,kHz are shown in \fref{fig:6}(a) and (c), respectively. Due to the slight radial anisotropy, both crystals are extended in the $yz$-plane, leading to a coupling of the $y$ and $z$ modes. 

The 10-ion crystal is axially $57$\,$\mu$m long and the two central ions (labelled as 5 and 6) are at a distance of $1.7$\,$\mu$m from the nodal line. The 10-ion crystal is point symmetric (\fref{fig:6}(a)), i.e.~it is symmetric with respect to the crystal center. We can identify pairs of ions, which are positioned opposite to each other at equal distance from the nodal line ($y_j = - y_{N+1-j}$ with N being the number of ions). These paired ions are subject to opposite gradients ($\vec{g}_j = -\vec{g}_{N+1-j}$), e.g.~the two central ions (5,6). The radial vibrational modes can be divided in two groups, those that have point symmetry and those that have mirror symmetry with respect to the $y$ axis (at $z=0$). A point-symmetric mode behaves like the oop mode of the radial two-ion crystal (\sref{two-ion-radial}), where the mode vectors point in opposite direction ($\beta_j^y  =  -\beta_{N+1-j}^y$) and the force contributions add constructively, leading to heating. The most heated mode is shown in the upper panel of \fref{fig:6}(b). Assuming the rf noise  $S_\mathrm{P} = 1.5\times10^{-14}$\,W/Hz in our system, it heats with $31.7$\,ph\,s$^{-1}$ ($0.3$\,mKs$^{-1}$) at a mode frequency of $\osec=2\pi\times185$\,kHz. It is the breathing-like mode in the $z$ direction and has the largest number of nodes in the $y$ direction. In contrast, in a mirror-symmetric mode, the mode vectors of two ions in a pair point in the same direction and have the same mode amplitude $\beta_j^y  =  \beta_{N+1-j}^y$. Similarly as for the com mode of the radial two-ion crystal, their force contributions cancel and the mode is protected from this source of heating. 
An example of a mirror-symmetric mode at $\osec=2\pi\times561$\,kHz is shown in the lower panel of \fref{fig:6}(b).

The 11-ion zig-zag crystal shown in \fref{fig:6}(c) is 60\,$\mu$m long and the central ion (6) is at 2.1\,$\mu$m from the nodal line. In contrast to the 10-ion crystal, it has mirror symmetry, leading to heating of mirror-symmetric modes, while point-symmetric modes are protected. This can be understood by drawing the analogy with the radially displaced \textit{axial} two-ion crystal (see \fref{fig:4}(a)). Again, symmetrically placed pairs of ions $y_j = y_{N+1-j}$ are identified, e.g.~ions (5,7) or (4,8) with $\vec{g}_j = \vec{g}_{N+1-j}$. 
But now, if the pairs of ions show a oop-like motion ($\vec{\beta}_{\alpha,j} = - \vec{\beta}_{\alpha,N+1-j}$), pairwise cancellation occurs and, consequently, point-symmetric modes are protected from heating (upper panel \fref{fig:6}(d)).
For mirror-symmetric modes (lower panel \fref{fig:6}(d)), the pairs of ions behave like a com mode ($\vec{\beta}_{\alpha,j} = \vec{\beta}_{\alpha,N+1-j}$), where the force contributions add up constructively and rf-noise induced heating occurs. For the 11-ion crystal, the highest heating rate is calculated to be $24.6$\,ph\,s$^{-1}$ ($0.2$\,mK\,s$^{-1}$) at $\osec=2\pi\times195$\,kHz. 
Note, that in the 11-ion crystal, the centre ion $j = (N+1)/2$ is unpaired. It has a finite radial mode amplitude in mirror-symmetric modes which are heated, while it is radially at rest in point-symmetric modes ($\beta^{x,y}_{\alpha,(N+1)/2} = -\beta^{x,y}_{\alpha,(N+1)/2} =0$), which are protected from heating. For most of the heated modes, there are also partial cancellations, e.g.~the force adds up pairwise, but partially cancels with respect to other ion pairs. For a full list of motional modes of the 11-ion crystal, see \ref{appendixC}.

\begin{figure}[phtb]
	\begin{center}

		\includegraphics[clip,width=\textwidth]{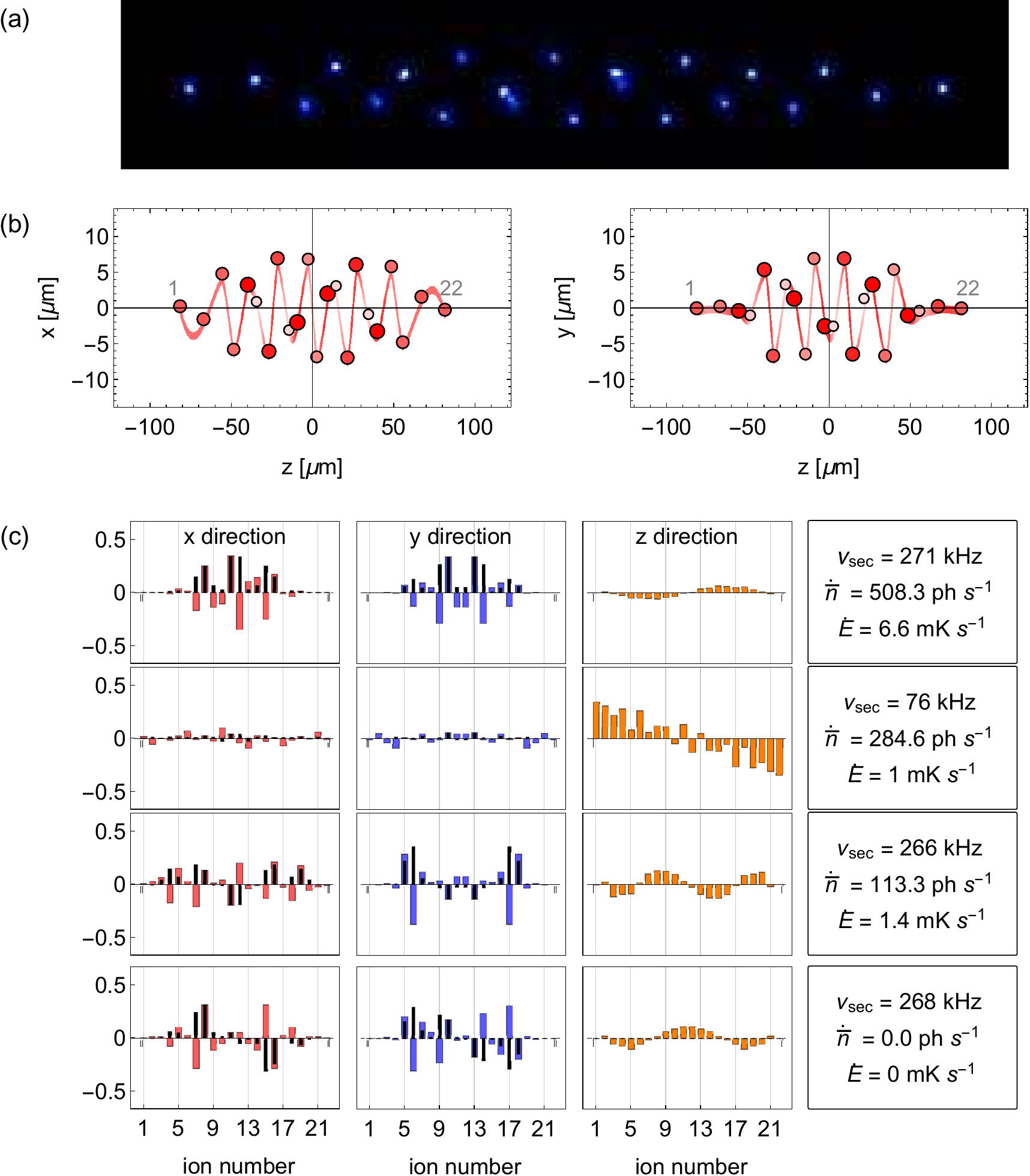}
	\end{center}
	\caption{Heating of a 3D extended crystal}{
		(a) Experimental image of a 22-ion helix crystal, with a length of 160\,$\mu$m and a radial diameter of 13\,$\mu$m.\\ 
		(b) Calculated ion positions in two planes ($xz$ and $yz$) for a 22-ion helix at similar confinement as in (a) at $(\omega_\mathrm{x},\omega_\mathrm{y},\omega_\mathrm{z}) = 2\pi\times(201,\,200,\,50)$\,kHz. To guide the eye, the position of the ion in the direction perpendicular to the plane is indicated by the size and shade of the points and the ions are connected by a spline. The crystal is point symmetric in the $xz$ plane and mirror symmetric in the $yz$ plane.\\
		(c) Panels with mode vector amplitudes in the $x$ (red bars), $y$ (blue) and $z$ (orange) direction of four exemplary modes together with force contributions (see text). The calculated heating rates, both in phonon number and energy, for each mode are shown together with the mode frequency in the textbox. In the most heated modes, ions that have a large radial displacement have a large mode amplitude. However, if the frequency of the mode is low, the $\dot{\bar{n}}$ becomes large and the mode is also significantly heated. \\
	}
	\label{fig:7}
\end{figure}
\subsection{3D crystals}\label{3Dcrystal}
As an example for a three dimensional crystal, we take a 22-ion helix \cite{Birkl:1992,Schiffer:1993,Dubin:1993}  at trap frequencies $(\omega_\mathrm{x},\omega_\mathrm{y},\omega_\mathrm{z}) = 2\pi\times(201,\,200,\,50)$\,kHz. It has a crystal length of 163\,$\mu$m and a maximum radial extension of 7\,$\mu$m. An experimentally obtained image of such a crystal under similar conditions is shown in \fref{fig:7}(a). The calculated ion positions in two planes are plotted in \fref{fig:7}(b), where the size and the brightness of the colour of the points indicate the position in the third dimension. 
The panels in \fref{fig:7}(c) show the normalized mode amplitudes and the individual force contributions in all three directions for four exemplary modes. Similar as in \sref{2Dcrystal}, the behaviour can be understood by looking at the symmetries of the crystal. In the $xz$ plane the crystal is point symmetric ($x_j = -x_{N+1-j}$, $g^x_j = -g^x_{N+1-j}$) and in the $yz$ plane the crystal is mirror-symmetric ($y_j = y_{N+1-j}$, $g^y_j = -g^y_{N+1-j}$). Consequently, modes with point-symmetric $x$ components and mirror-symmetric $y$ components are heated. The upper panel of \fref{fig:7}(c) shows the most heated mode at $\osec=2\pi\times271$\,kHz with a heating rate of $\dot{\bar{n}} = 508$\,ph\,s$^{-1}$ (6.6\,mK\,s$^{-1}$), where the biggest contribution comes from a few ions with large mode amplitudes, located in the helix region. The second panel shows a mode with much smaller and evenly distributed mode amplitudes, both in the $x$ and $y$ direction, which is still significantly heated (1.0\,mK\,s$^{-1}$) due to the low mode frequency and, therefore, a big change in phonon number of $\dot{\bar{n}} = 285$\,ph\,s$^{-1}$.
In addition, due to the transmission function of the resonant circuit, the noise gets amplified much more at lower secular frequencies. The lower two panels show two modes which are very similar, but have the opposite symmetry, both in the $x$ and $y$ direction, leading to rf-noise induced heating in one case and full cancellation in the other.  

Summing over all modes, we obtain a total heating from rf noise of this crystal of $1.57\times10^{-25}$\,J\,s$^{-1}$ corresponding to $\dot{E}/k_\mathrm{B}=11.4$\,mK s$^{-1}$, of which 58\% is contributed by a single mode (the mode shown in the upper panel in \fref{fig:7}(c)). Due to heating of individual modes, the crystal is not in thermal equilibrium though. For comparison, we estimate the heating of the crystal due to DC noise taking into account the scaling with frequency and ion number. Here, only the three com modes are heated by 1.5\,mK\,s$^{-1}$, 1.5\,mK\,s$^{-1}$ and 6.1\,mK\,s$^{-1}$ and the total energy increase is $1.26\times 10^{-25}$\,J\,s$^{-1}$. The low-frequency com mode in the $z$ direction shows the largest heating rate due to the frequency scaling of the DC electronic noise, see \fref{fig:1}(a).

\section{Implications on spectroscopy}
Extended ion crystals are promising candidates for more stable optical clock operation \cite{Champenois:2010, Arnold:2015,Herschbach:2012,Keller:2019b} and for quantum simulation, which are rapidly advancing towards higher dimensional crystals \cite{Zhang:2017,Richerme:2016}. 
While the approach of arrays of 1D crystals, with tens of ions each, allows for a high level of control of systematic shifts at the $10^{-19}$ level \cite{Herschbach:2012,Keller:2019b}, clock spectroscopy in 3D crystals was proposed in \cite{Arnold:2015} in view of the possible larger number of ions, ranging up to 1000 ions in a single crystal. The latter approach possibly will require cryogenic setups to handle the collisional heating of large Coulomb crystals, but also will require a careful investigation of the complex many-body dynamics of a large Coulomb crystal with $3\times N$ motional modes (= phonons).

Here, we briefly discuss the consequences of rf-noise induced heating of ion Coulomb crystals on spectroscopy and coherent manipulation using the 3D crystal of 22 $^{172}$Yb$^{+}$ ions studied in \sref{3Dcrystal} as an example. For comparison, we first estimate second-order Doppler shifts due to the simple presence of the rf trapping field in the 3D crystal configuration. We then calculate the second-order Doppler shifts due to heating of motional modes via rf noise. Finally, we estimate the reduction and the fluctuation of the Rabi frequency, for individual ions inside the Coulomb crystal due to a finite thermal excitation of motional modes. 
Since the Rabi-frequency governs the excitation probability of the ions, spatial and temporal variations will limit the stability of a clock measurement.
We consider both the residual excitation of modes after laser cooling, and the additional heating of motional modes of the 3D crystal during the clock interrogation time.

\subsection{Second-order Doppler shift due to excess micromotion}
The relative second-order Doppler shift (time dilation shift) due to micromotion can be calculated from $E_\mathrm{rf}$ at the ion position \begin{equation}\label{eq:timedilation}
	\left<\frac{\Delta \nu_\mathrm{td, j}}{\nu_0}\right> = - \left(\frac{eE_\mathrm{rf}(\vec{r}_j)}{2mc\Orf} \right)^2
\end{equation}
where $\nu_0$ is the transition frequency, $e$ the electron charge, and $c$ is the speed of light.
For the 22-ion crystal, the calculated relative shifts are all below $\Delta \nu_\mathrm{td}/\nu_0 =- 4.9\times 10^{-16}$. The highest values are obtained for ions $j = 11$ and $j=12$, which have the largest radial extension in the helix region at $E_\mathrm{rf} = 3.6$\,Vmm$^{-1}$ (7\,$\mu$m radial distance from nodal line). For the inner 18 ions, forming the helix, the shift ranges between $-2\times 10^{-16}$ and $-4.9\times 10^{-16}$. The use of specific ion species with a negative differential static scalar polarizability $\Delta \alpha_0$, enables the cancellation of two rf induced shifts. By operating at the `magic' rf drive frequency, the time dilation shift and the scalar AC Stark shift can cancel each other \cite{Berkeland:1998,Barrett:2015,Dube:2014,Huang:2016}.

\subsection{Thermal second-order Doppler shift}
In a crystal of ions at finite temperature, the thermal motion of the ions within the vibrational modes induces a time dilation shift according to the mean squared ion velocities $\left<v_j^2\right>$. These are obtained from the mode amplitudes $\vec{\beta}_{\alpha,j}$ and the mode temperature $T_\alpha=\bar{n}_\alpha\hbar\omega_\alpha/k_\mathrm{B}$
\begin{equation}
\left<\frac{\Delta \nu}{\nu_0}\right>_j =- \frac{\left<v_j^2\right>}{2c^2} =-\frac{1}{2mc^2} \sum_{\alpha=1}^{3N}k_\mathrm{B}T_\alpha\vec{\beta}_{\alpha,j}^2
\end{equation}
For the 22-ion crystal of $^{172}$Yb$^+$ cooled to Doppler temperature ($T_\alpha= T_\mathrm{D}$ = 0.5\,mK), the motional state occupation of, e.g.,~the radial and axial com mode are $\bar{n} =52$\,ph and $\bar{n} =208$\,ph, respectively. 
Assuming that the crystal is in thermal equilibrium, the relative time dilation shift for each of the ions is $-4\times 10^{-19}$. At a crystal temperature of $T=5\,\mu$K, this shift reduces by two orders of magnitude to $-0.04\times10^{-19}$.

Due to heating by rf fields in the extended crystal, we obtain an increase in time dilation shift for ion $j$, according to
\begin{equation}
\frac{\partial}{\partial t}\left<\frac{\Delta \nu}{\nu_0}\right>_j =- \frac{1}{2mc^2}\sum_{\alpha=1}^{3N}\dot{\bar{n}}_\alpha\hbar\omega_\alpha\vec{\beta}_{\alpha,j}^2.
\end{equation}
This increase is inhomogeneous over the 22-ion crystal due to heating of particular modes only, and ranges between $-0.7 \times 10^{-19}$ and $-3 \times 10^{-19}$ per second for the individual ions, in the case of a quality factor $Q= 542$ of the resonant circuit. This increase would be higher by a factor of 2.4 for reduced quality factor of $Q=204$. Note that this dynamic shift can not be suppressed by using a magic drive frequency.
\subsection{Debye-Waller-effect}
The thermal motion of the ions also affects the coupling between the ions and light fields. Firstly, the motion along the propagation direction of the interrogation laser reduces the average Rabi frequency $\bar{\Omega}_j$ of the ion $j$ relative to the free atom Rabi frequency $\Omega^0$ \cite{Wineland:1998}
\begin{equation}\label{eq:DWrabi}
\frac{\overline{\Omega}_j}{\Omega^0} = \prod_{\alpha=1}^{3N} \exp{\left[-\eta_{\alpha,j}^2 (\bar{n}_\alpha + 1/2)\right]}.
\end{equation}
where $\eta_{\alpha,j}$ is the Lamb-Dicke-factor of mode $\alpha$ for the $j$th ion.
Secondly, it introduces a fluctuation $\sigma_{\Omega_j}$ of the Rabi frequency between subsequent experiments. The relative scatter from shot to shot (rms) can be calculated using modified zeroth-order Bessel functions $I_0(x)$ \cite{Wineland:1998}
\begin{equation}\label{eq:DWrabiFluct}
\sigma_{\Omega_j}=\frac{\Delta\Omega^{(rms)}_j}{\overline{\Omega}_j}\ = \sqrt{\left[\prod_\alpha I_0\left(2\eta_{\alpha,j}^2\sqrt{\bar{n}_\alpha (\bar{n}_\alpha +1)}\right)\right]-1}.
\end{equation}
For radial interrogation on the electric quadrupole transition $\mathrm{^2S_{1/2}}$$\rightarrow$$\mathrm{^2D_{5/2}}$ at 411\,nm (see \fref{fig:1}) at the Doppler temperature ($\bar{n} =52$\,ph), the Rabi frequencies for all ions are reduced to $\overline{\Omega}/\Omega^0\lesssim9\,\%$. For axial interrogation ($\bar{n} =208$\,ph), they are all $\overline{\Omega}/\Omega^0\lesssim1\,\%$ and the shot-to-shot scatter is about 100\,\% of the reduced Rabi frequency in both cases.
 
Even at ideal conditions of a crystal cooled to close to the motional ground state, the effect remains significant. At a crystal temperature of 5\,$\mu$K, the mean occupation number of the radial com modes is $\bar{n} \lesssim 0.3$\,ph and that of the axial com mode is at $\bar{n} \approx 1$\,ph. For radial interrogation, the Rabi frequencies range between 11\,\% ($\sigma_{\Omega_j} \approx 38\,\%$) and 96\,\% ($\sigma_{\Omega_j}\approx 2\,\%$) for the most affected and the least affected ion, respectively. A more favourable direction of interrogation is along the crystal axis. In this case, the Rabi frequencies range between 83\% and 92\% of the free atom Rabi frequency, and the shot-to-shot fluctuations are $\sigma_\mathrm{rms}\lesssim5\,\%$. Within 100\,ms of rf-noise induced heating, the distribution becomes less homogeneous, as the Rabi frequencies reduce to between 54\% and 75\%, with a shot-to-shot noise between $\sigma_{\Omega_j}\approx 18\,\%$ and 12\,\%, respectively, through coupling between radial and axial modes. Note that in a planar 2D radial crystal, the coupling of radial heating to axial modes can be strongly reduced\cite{Richerme:2016}.
\FloatBarrier
\section{Conclusion}\label{conclusion}
We have experimentally studied heating effects from electric field noise on the motion of trapped ions, originating from both dc and rf electric fields. The previously reported low heating rate of about $0.7$\,ph\,s$^{-1}$ per ion at $\osec = 2\pi\times620$\,kHz due to noise of static electric fields \cite{Keller:2019b} was not significantly changed over a time period of four years and could be confirmed for linear chains of up to 11 ions. The heating rate of the first out-of-phase mode of a $100\,\mu$m-crystal was below our measurement resolution of 0.2\,ph\,s$^{-1}$. The electric field noise spectral density ($S_\mathrm{E}\left(\nu\right)=8.49(8)\times10^{-9} \mathrm{(V/m)}^2/\nu$) is in the lower range of values reported for other traps\cite{Brownnutt:2015}.

We have measured heating rates in the presence of excess micromotion by radially displacing a single ion and investigated the influence of the quality factor Q of the resonant circuit by reducing it from $Q=542$ to $Q=204$. This led to a significant increase of motional heating of up to a factor 2.8 depending on the secular frequency. For a two-ion crystal, heating of the com mode is observed, while the out-of-phase mode is protected from rf-noise induced heating, even at a large radial displacement of $\Delta r = 21$\,$\mu$m, confirming the suppression of this heating mechanism in the out-of-phase modes in linear ion chains.

Based on our experimental result and measured voltage noise spectral density, we have calculated the motional heating effect of rf noise on radially extended 2D and 3D crystals. The rf-noise induced heating of any mode is largely determined by the symmetry of the crystal and of the mode. Ions which are far away from the nodal line and have a large mode amplitude contribute most. In a 3D 22-ion helix-shaped crystal, the energy increases by $\dot{E}/k_\mathrm{B}=11.4$\,mK s$^{-1}$ due to rf noise dominated by a single mode (6.6\,mK\,s$^{-1}$) and is comparable to heating of com modes induced by dc electric field noise (1.5\,mK\,s$^{-1}$ for the radial modes and 6.1\,mK\,s$^{-1}$ for the axial mode). 

The micromotion-induced fractional second-order Doppler shift $\Delta \nu/\nu_0$ is on the level of $10^{-16}$, which can be cancelled in some specific ion species by use of a magic trap rf drive frequency \cite{Berkeland:1998,Barrett:2015,Dube:2014,Huang:2016}.
The calculated rf-noise induced heating leads to an increase of the second-order Doppler shift of up to $-3\times10^{-19}$\,s$^{-1}$, which is inhomogeneous across the crystal and cannot be suppressed by a magic frequency drive. By reducing the quality factor of the rf circuit from $Q=542$ to $Q=204$, this shift increases by a factor of 2.4. Furthermore, the Debye-Waller effect leads to a dominating noise contribution to excitation of the atomic qubit. Therefore, the exact crystal configuration should be carefully chosen and ground state cooling is critical. The calculated spectroscopic properties and heating rates are highly dependent on the quality factor $Q$ of the resonant circuit. Strong filtering of rf noise is required for precision spectroscopy of 2D and 3D crystals, such as is proposed for multi-ion clock-based experiments \cite{Arnold:2015} and scalable quantum information processing \cite{Zhang:2017,Richerme:2016}.

\ack 
We gratefully thank J Keller, I Vybornyi, C Roos and K Hammerer for fruitful discussions and R Nigmatullin for providing parts of the calculation code. L\,S\,D acknowledges support from the Alexander von Humboldt foundation. This project has been funded by the Deutsche Forschungsgemeinschaft (DFG, German Research Foundation) under Germany’s Excellence Strategy – EXC-2123 QuantumFrontiers – 390837967 (RU B06) and through Grant No. CRC 1227 (DQ-mat, project B03). This project 17FUN07 CC4C has received funding from the EMPIR programme co-financed by the Participating States and from the European Union's Horizon 2020 research and innovation programme.
\FloatBarrier

\appendix
\section{Temperature determination of larger ion crystals}\label{appendixA}
\begin{figure}[t]
	\begin{center}

		\includegraphics[clip,width=\textwidth]{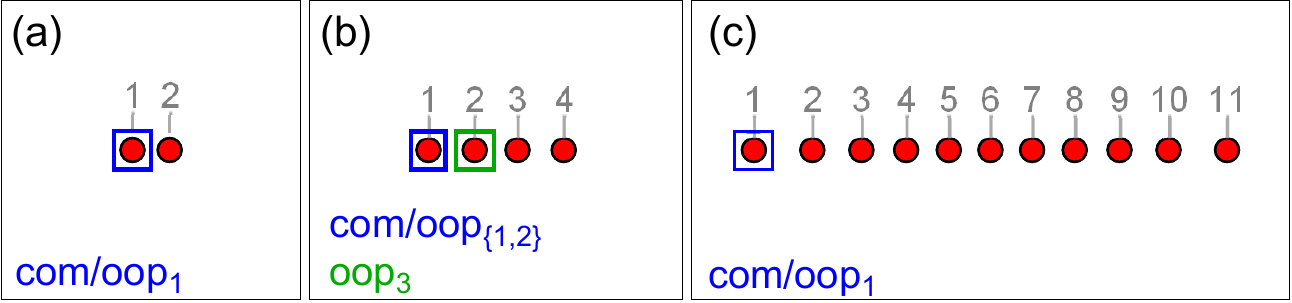}
	\end{center}
	\caption{Ion selection for state-detection in heating-rate measurements}{The ions are numbered from left to right as seen on the EMCCD camera in the experiment.\\
		(a) Ion 1 (blue square) was selected for heating rate measurements of all modes in the two-ion crystal. \\
		(b) In the four-ion crystal, ion 1 (blue square) was selected for heating rate measurements of the com and the first and second oop mode. For heating rate measurements of the fourth oop mode, the second ion (green square) was selected for better coupling.\\
		(c) In the 11-ion crystal, ion 1 (blue square) was selected for heating rate measurements of the com and the first oop mode.
	}
	\label{fig:A1}
\end{figure}
The heating rate measurements of larger ion chains were done with global addressing and single-ion state-detection. The ion selected for detection in the three measured crystals are shown in \fref{fig:A1}. Global addressing of ions on motional sidebands creates entangled states, which has to be taken into account for accurate temperature determination using resolved sideband thermometry. In this section, we elaborate on this effect and its influence on the measurements. For simplicity, only the derivation for a two-ion crystal is given.

The two involved electronic levels are denoted by $\ket{\mathrm{^2S_{1/2}}}\equiv \ket{\downarrow}$ and $\ket{\mathrm{^2D_{5/2}}}\equiv\ket{\uparrow}$. The motional state of the centre-of-mass (com) mode at frequency $\omega_\mathrm{c}$ is indicated with quantum number $n$ and that of the out-of-phase (oop) mode at $\omega_\mathrm{s}$ with quantum number $k$. For simplicity, we will only consider motional excitation of the com mode. The Lamb-Dicke parameter of the com mode for two ions is given by $\eta_\mathrm{c} = \eta / \sqrt{2}$, where $\eta = \frac{\lambda}{2\pi} \sqrt{\hbar/2m\omega_\mathrm{c}}$ and the Rabi frequency of the carrier, first red- and blue sideband are $\Omega$, $\sqrt{n} \eta_\mathrm{c} \Omega$ and $\sqrt{n+1} \eta_\mathrm{c} \Omega$, respectively. The interaction Hamiltonian for two ions that are both addressed with equal intensity is given by (neglecting terms of order $\eta^2$)
\begin{equation}
\label{Hamiltonian_two_ion}
\eqalign{
\fl H_\mathrm{I} &= \sum_{l=1}^2 \frac{\hbar \Omega}{2}e^{-i(\delta t -\phi_l)}\ket{\downarrow_l}\bra{\uparrow_l}\Big[\ket{n,k}\bra{n,k}+i\eta_\mathrm{c} (\sqrt{n}\ket{n-1,k}\bra{n,k}e^{-i\omega_\mathrm{c} t}\\
\fl&+\sqrt{n+1}\ket{n+1,k}\bra{n,k}e^{i\omega_\mathrm{c} t}) \Big]+ h.c.,}
\end{equation}
where $l$ indicates index of the ion and $\phi_l$ is the phase of the laser at ion $l$. If only resonant terms are taken into account, the Hamiltonian for the first red- and blue sideband are given by
\begin{eqnarray}
\fl H_\mathrm{r} = i \eta_\mathrm{c} \sqrt{n} \frac{\hbar \Omega}{2}\ket{n,k}\bra{n-1,k}\Big[\ket{\downarrow_1}\bra{\uparrow_1}e^{i\phi_1}+\ket{\downarrow_2}\bra{\uparrow_2}e^{i\phi_2}\Big] +h.c.,\\
\fl H_\mathrm{b} = i \eta_\mathrm{c} \sqrt{n+1} \frac{\hbar \Omega}{2}\ket{n,k}\bra{n+1,k}\Big[\ket{\downarrow_1}\bra{\uparrow_1}e^{i\phi_1}+\ket{\downarrow_2}\bra{\uparrow_2}e^{i\phi_2}\Big] +h.c.,
\end{eqnarray}
respectively. 

In the experiment, both ions are first prepared in the electronic ground state  $\ket{\downarrow \downarrow,n}$. By tuning the frequency to the first red sideband, the ions can be in one of four possible final states $\ket{\downarrow \downarrow,n}$, $\ket{\downarrow \uparrow,n-1}$, $\ket{\uparrow \downarrow,n-1}$ and $\ket{\uparrow \uparrow,n-2}$ taking either 0, 1 or 2 phonons out of the system. Similarly, the four possible final states of the blue sideband are $\ket{\downarrow \downarrow,n}$, $\ket{\downarrow \uparrow,n+1}$, $\ket{\uparrow \downarrow,n+1}$ or $\ket{\uparrow \uparrow,n+2}$.\\
The time dependent evolution of the state can be found by using the propagator $U = V e^{-iV^{-1}HVt/\hbar}V^{-1}$, which yields \cite{Home:2000,King:1998}
\begin{eqnarray}
\label{eq:RSB_twoion}
\eqalign{
\fl \ket{\Psi_\mathrm{r}} = &e^{-i(\phi_1+\phi_2)}\frac{\sqrt{n(n-1)}}{2n-1}[1-\cos(g_\mathrm{r}\Omega t)]\ket{\uparrow \uparrow,n-2}\\ 
\fl &+ e^{-i\phi_2}\sqrt{\frac{n}{2(2n-1)}}\sin(g_\mathrm{r}\Omega t)\ket{\uparrow \downarrow,n-1}\\
\fl &+e^{-i \phi_1}\sqrt{\frac{n}{2(2n-1)}}\sin(g_\mathrm{r}\Omega t)\ket{\downarrow \uparrow,n-1}+\Big\{1-\frac{n}{2n-1}[1-\cos(g_\mathrm{r} \Omega t)]\Big\}\ket{\downarrow\downarrow,n},}\\
\label{eq:BSB_twoion}
\eqalign{
\fl \ket{\Psi_\mathrm{b}} = &e^{-i(\phi_1+\phi_2)}\frac{\sqrt{(n+1)(n+2)}}{2n+3}[1-\cos(g_\mathrm{b}\Omega t)]\ket{\uparrow \uparrow,n+2}\\
\fl &+e^{-i\phi_2}\sqrt{\frac{n+1}{2(2n+3)}}\sin(g_\mathrm{b}\Omega t)\ket{\uparrow \downarrow,n+1}\\
\fl &+e^{-i \phi_1}\sqrt{\frac{n+1}{2(2n+3)}}\sin(g_\mathrm{b}\Omega t)\ket{\downarrow \uparrow,n+1}+\Big\{1+\frac{n+1}{2n+3}[1-\cos(g_\mathrm{b} \Omega t)]\Big\}\ket{\downarrow\downarrow,n},}
\end{eqnarray}
where $g_\mathrm{r} = \eta_\mathrm{c}\sqrt{(2n-1)/2}$ and $g_\mathrm{b} = \eta_\mathrm{c}\sqrt{(2n+3)/2}$.\\
In the experiment, only the state of one of the two ions was detected. Therefore, the excitation probability of the red- and blue sideband are given by $|\braket{\uparrow,n'|\Psi_\mathrm{r}}|^2 = \braket{\uparrow \downarrow,n-1|\Psi_\mathrm{r}}^2+|\braket{\uparrow \uparrow,n-2|\Psi_\mathrm{r}}|^2$ and $|\braket{\uparrow,n'|\Psi_\mathrm{b}}|^2 = \braket{\uparrow \downarrow,n+1|\Psi_\mathrm{b}}^2+|\braket{\uparrow \uparrow,n+2|\Psi_\mathrm{b}}|^2$, respectively. Using \eref{eq:RSB_twoion} and \eref{eq:BSB_twoion} and a thermal Maxwell-Boltzmann state distribution $p(n)$ the red and blue sideband amplitude can be written as
\begin{eqnarray}
\label{eq:ampRSB}
\eqalign{
\fl p_{\mathrm{r},\uparrow}&= \sum_{n=0}^{\infty} p(n+2) \frac{n(n-1)}{(2n-1)^2}[1-\cos(g_\mathrm{r}\Omega t/2)]^2\\
\fl&+p(n+1)\frac{n}{2(2n-1)}\sin^2(g_\mathrm{r} \Omega t)}\\
\label{eq:ampBSB}
\eqalign{
\fl p_{\mathrm{b},\uparrow}&= \sum_{n=0}^{\infty} p(n) \frac{(n+1)(n+2)}{(2n+3)^2}[1-\cos(g_\mathrm{b}\Omega t/2)]^2\\
\fl &+ p(n)\frac{n+1}{2(2n+3)}\sin^2(g_\mathrm{b}\Omega t)}
\end{eqnarray}
\begin{figure}[t]
	\begin{center}
		\includegraphics[clip,width=\textwidth]{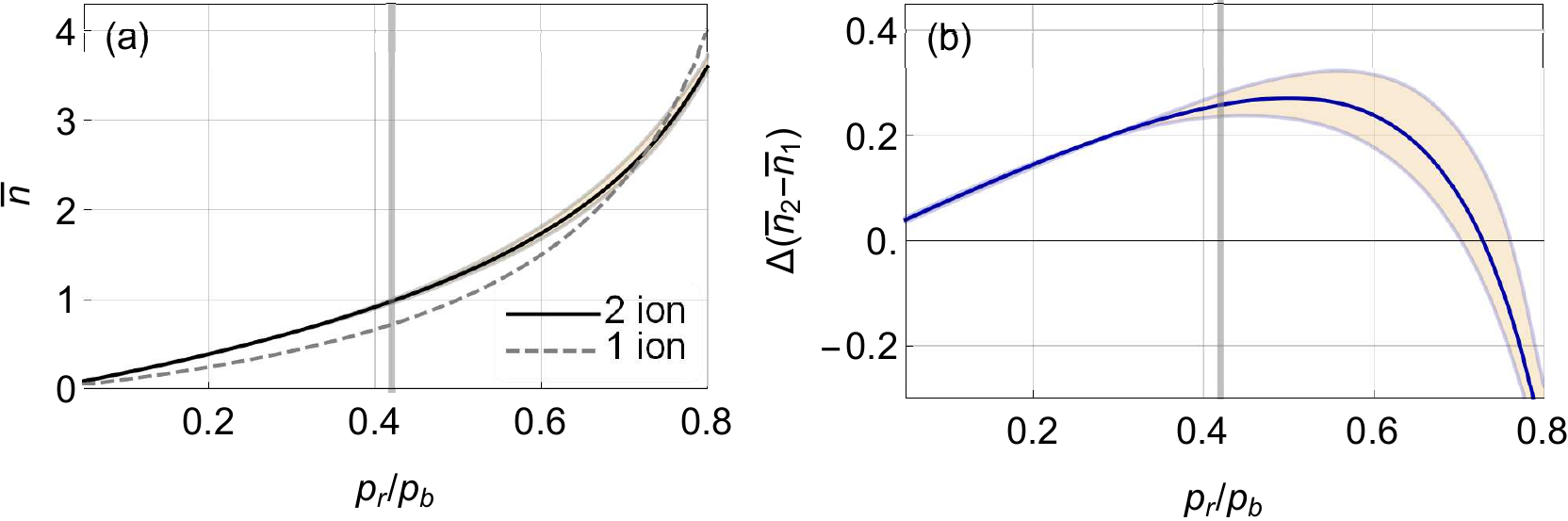}
	\end{center}
	\caption{Temperature as a function of the red and blue sideband ratio for a two-ion crystal.}{
		(a) Mean occupation number as a function of the red- and blue side ratio obtained from the numerical solution of the two-ion wavefunction (black curve) at optimal pi-time. The shaded area indicates the deviation from having a 10\% different pulse time. For comparison, the curve of a single-ion (dashed line) is shown. \\
		(b) Difference between the two-ion result and the single-ion relation. The shaded area indicates the deviation from having a 10\% different pulse time. \\
		The vertical grey line in (b) and (c) indicates the limit of $\bar{n}=1$ below which we typically operate.
	}
\label{fig:A2}
\end{figure}
From \eref{eq:ampRSB} and \eref{eq:ampBSB}, $\bar{n}$ as a function of sideband ratio can be obtained numerically for a specific interrogation time $t$, as can be seen in \fref{fig:A2}(a). Similar as in the experiment, the interrogation time is fixed to reach maximum excitation probability on the blue sideband directly after ground state cooling, when a typical temperature of $\bar{n}=0.05$ is reached. However, the determined temperature is no longer independent of the interrogation time, as it is for a single ion. The deviation from having a 10\% longer or shorter interrogation pulse is shown by the shaded area in \fref{fig:A1}. For comparison also the curve for a single-ion, as given by \eref{eq:n}, is shown with the dashed line in \fref{fig:A2}. For our operating range, $\bar{n}\leq1$ (grey vertical line in \fref{fig:A2}), the two-ion result is systematically higher than that of a single ion. In this range, it can be well approximated by a linear function, where the slope of the two-ion result is 35\% larger than that of a single ion. The absolute difference between the two curves is shown in \fref{fig:A2}(b). The largest deviation occurs at $p_\mathrm{r}/p_\mathrm{b}=0.5$ and is $\Delta\bar{n}=0.27$ ph. The shaded area indicates the calculated value with a $\pm10\%$ different interrogation time. 

The results show that when the state of only a single ion is detected, while sidebands are globally driven, a systematically too low temperature is obtained from the simple relation given by \eref{eq:n}. In order to correct for this, the experimental data is scaled by a factor 1.35(4). The additional uncertainty on this factor stems from taking into account a $\pm 10\%$ uncertainty of the pulse area. Note that the correction factor of the oop mode is nearly equal to that of the com mode, because the normalized mode amplitude of the detected ion remains the same for both modes and the mode frequencies are almost equal.

By extending the Hamiltonian of \eref{Hamiltonian_two_ion}, the same calculation can be carried out numerically to obtain correction factors for the com mode of the linear 4- and 11-ion linear chains. It yields correction factors of $2.68(4)$ and $7.05(5)$ for the com mode of 4 and 11 ions, respectively. The correction factors of all three oop modes of the linear 4-ion crystal were calculated by taking into account the ion-dependent Lamb-Dicke factors. The normalized mode amplitudes of the ions in 2$^{\mathrm{nd}}$ oop mode are the same as those in the com mode, leading to a nearly equal correction factor of $2.68(4)$. Similarly, due to the choice of ion in the experiment (see \fref{fig:A1}), the correction factor of $1.67(4)$ for the 1$^{\mathrm{st}}$ and 3$^{\mathrm{rd}}$ oop mode were nearly equal.
	
For the 11-ion oop mode, no correction factor was calculated due to its complexity. Instead, the error bar was increased to $0.2$ ph\,s$^{-1}$, which is a relative error of $150$\%. We are confident that the heating rate of this mode is well within this range, as the calculated correction factors for the heating rate \textit{per ion} of the other measured modes lead to a reduction of the measured values by $20$\% to $60$\%. Moreover, the other data points were all scaled down by their respective correction factor, thus the uncorrected heating rate obtained for the first oop mode of the 11-ion crystal is an overestimation. 

\FloatBarrier
\clearpage
\section{Influence on heating rates due to non-linear mode coupling}\label{appendixB}
	Heating due to non-linear coupling between the ground state (GS) cooled modes and Doppler cooled modes can be addressed by performing higher-order expansions of the mutual Coulomb interaction~\cite{Marquet:2003,Roos:2008,Nie:2009,Kiethe:2020} or using molecular dynamics simulations. Since we are treating large Coulomb systems with up to 22 ions, we decided to perform molecular dynamics simulations involving the full Coulomb potentials as described in the following. By determining the time evolution of ion motion, we verify that under the experimental conditions, there is no significant energy flow to the cooled modes from any other mode.
	We use the respective experimental trapping parameters for the 2, 4 and 11-ion linear crystal configurations investigated in \sref{dc-two}.
We initially sample all mode velocities from a Maxwell-Boltzmann distribution at around 1\,mK. To simulate GS cooling, we set the initial mode-velocities of the GS cooled modes to $0\,\mathrm{m}\,\mathrm{s}^{-1}$.
We propagate the ion crystals in time and deduce the average kinetic energy of the normal modes every 50\,ms by integration of their Fourier spectra, using a time sample of $T_\mathrm{fft}=0.46\,$ms with $n_\mathrm{fft}= 32768$ steps. Details about the method can be found in \cite{Fuerst:2018}. As the strength of the non-linear coupling depends on the mode energies of the Doppler-cooled modes, we average over 20 simulation runs for each crystal size using different initial conditions in each run.

\begin{figure}[hbt]
	\begin{center}

		\includegraphics[clip,width=\textwidth]{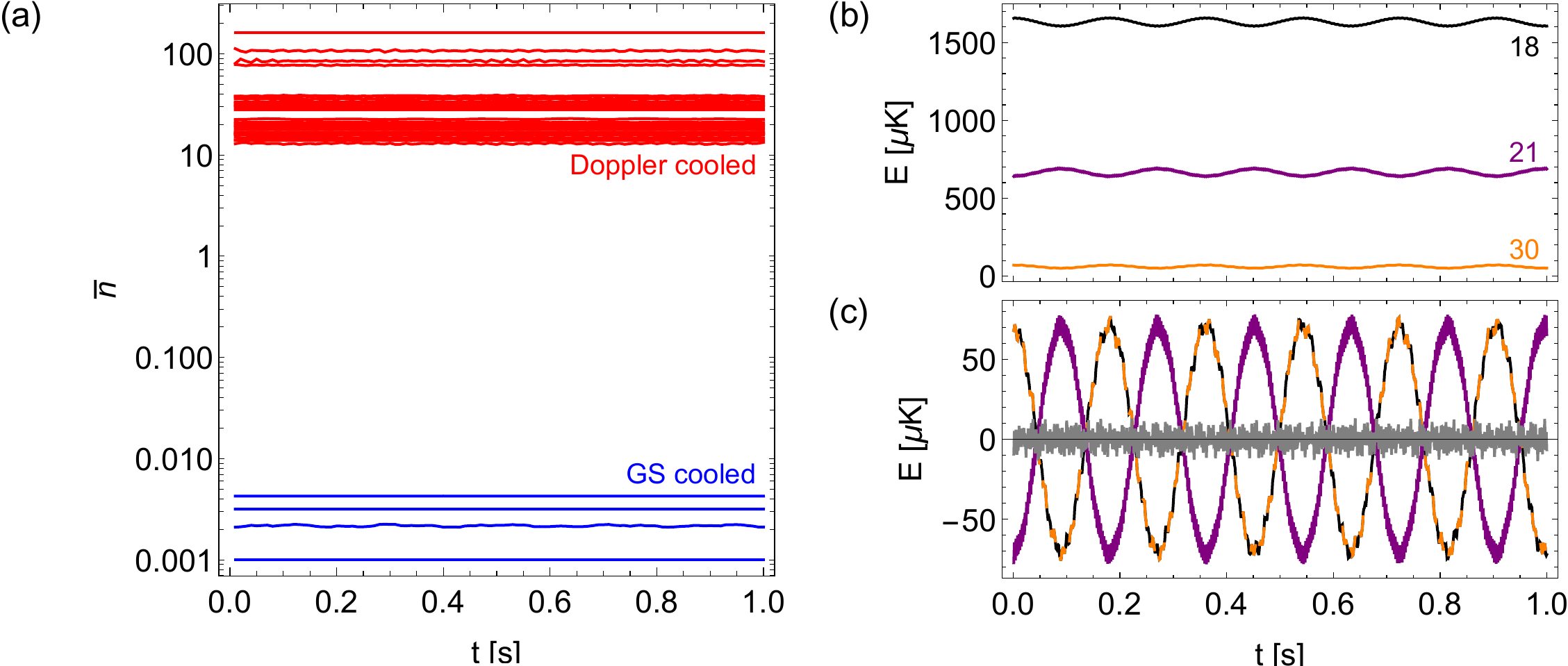}
	\end{center}
	\caption{Non-linear coupling between motional modes in an 11-ion crystal.}{
		(a) Mode occupation numbers over time for an 11-ion crystal at $T=1$\,mK (red lines), where the four highest-frequency radial modes (modes 1 through 4) along the $y$ direction have been set to $0\,\mathrm{m}\mathrm{s}^{-1}$ initial velocity (blue lines). For better visibility, $10^{-3}$\,ph is added to $\bar{n}$ of the ground state (GS) cooled modes to shift them upwards. Each line represents an average over 20 simulation runs.\\
		(b) Individual simulation run of mode temperatures is shown for three of the Doppler-cooled modes from (a) as an example of non-linear mode coupling in the 11-ion crystal (see text).\\
		(c) Amplitudes of energy oscillations during mode coupling. The energy of mode 18 and of mode 30 are added and multiplied by 2 (black-orange dashed line), the energy of mode 21 is multiplied by 3 (solid purple line), according to \eref{eq:modecoupling}. For each curve, the mean of the curve was subtracted to center them around zero. To indicate energy conservation, the sum of the two curves is also shown (grey line).
	}
	\label{fig:B1}
\end{figure}

For both the two-ion crystal and the four-ion crystal, where all radial modes in the directions with the strongest confinement ($y$) are cooled, we find no significant coupling between the GS cooled modes and Doppler cooled modes. In the case of the 11-ion crystal, only the four highest-frequency radial modes in $y$ direction were cooled. \Fref{fig:B1}(a) shows the calculated mean motional state occupation $\bar{n}$ of all motional modes over a period of a second. For the Doppler-cooled modes (red lines), the mean motional state occupation ranges between $\bar{n}=12$\,ph and $\bar{n}=166$\,ph. For better visibility, the plots of the four GS cooled modes (blue lines) are shifted upwards by $10^{-3}$\,ph. During 1 second, the mean motional state occupation of these four modes varies by $\Delta \bar{n}< 10^{-4}$\,ph.

We see non-linear coupling between some of the Doppler cooled modes, especially in the low frequency range. As an example, \fref{fig:B1}(b) shows the calculated mode energies  as function of time for a single simulation run. For clarity, only the energy of the three coupled modes are depicted. Two of them are radial modes at $\osecind{18}=2\pi\times526$\,kHz (black line) and at $\osecind{21}=2\pi\times485$\,kHz (purple)  along the weaker confined direction $x$  and the third mode is an axial mode at $\osecind{30}=2\pi\times200$\,kHz (orange). The index identifies the mode, sorted from highest (1) to lowest (33) mode frequency. Periodic energy exchange between the three modes is observed, where energy is transferred from mode 18 and 30, oscillating in phase with each other, to mode 21, oscillating out-of-phase. In this case, two phonons of both mode 18 and mode 30 are converted resonantly into three phonons of mode 21 and reversely,
\begin{equation}\label{eq:modecoupling}
	2\hbar \osecind{18} + 2\hbar\osecind{30} = 3\hbar\osecind{21}.
\end{equation}
This is illustrated in \fref{fig:B1}(c), where the curves from (b) are centred around zero by subtracting the mean mode energy. The energies of mode 18 and 30 are summed and multiplied by 2 (black-orange dashed line) according to \eref{eq:modecoupling}, while the energy of mode 21 is multiplied by 3 (purple line). The sum of the amplitudes (grey line) fluctuates around zero indicating energy conservation. The amount of energy exchanged in this process corresponds to two phonons at $\osecind{21}=2\pi\times485$\,kHz, which can be relevant, if heating rates of the weaker confined radial direction ($x$) or of the axial modes are considered. Note that we only measured heating rates along the $y$ direction and were not bothered by this effect. In addition, a slight change of confinement shifts the secular frequencies away from this resonance \cite{Marquet:2003}.

\clearpage\FloatBarrier
\section{All modes of the 11-ion crystal}\label{appendixC}
\begin{figure}[htbp]
	\begin{center}

		\includegraphics[clip,width=0.78\textwidth]{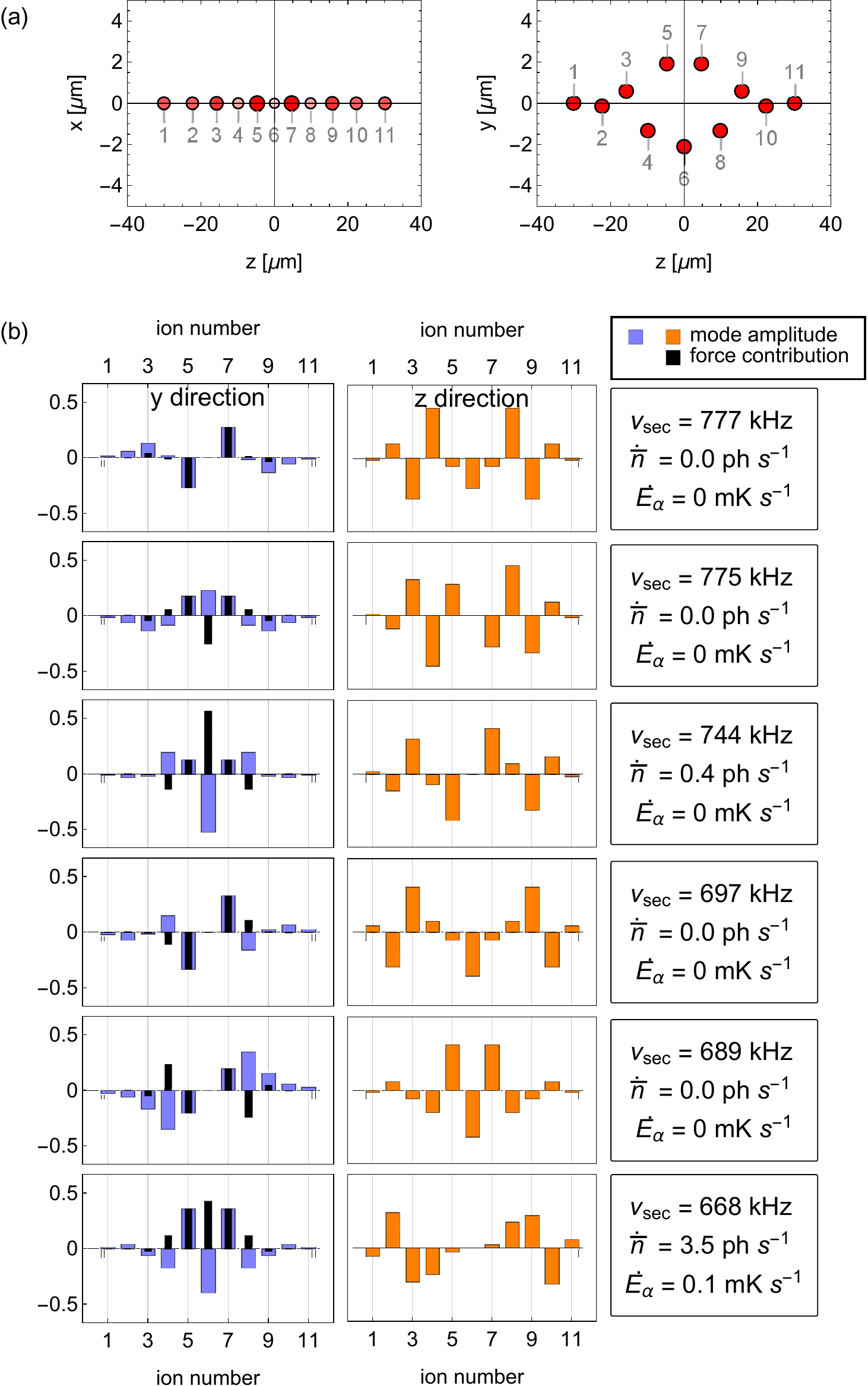}
	\end{center}
	\caption{RF-noise induced heating of a 2D 11-ion crystal.}{
		(a) Crystal geometry of the zig-zag crystal in two directions.\\
		(b) Mode amplitude of the individual ions in two directions and the corresponding force contributions. The frequency of the mode and the calculated heating rates are given both in phonon number and energy transfer in the text box on the right.
		}
\label{fig:C1}
\end{figure}
\begin{figure}[h]
	\begin{center}

		\includegraphics[clip,width=0.8\textwidth]{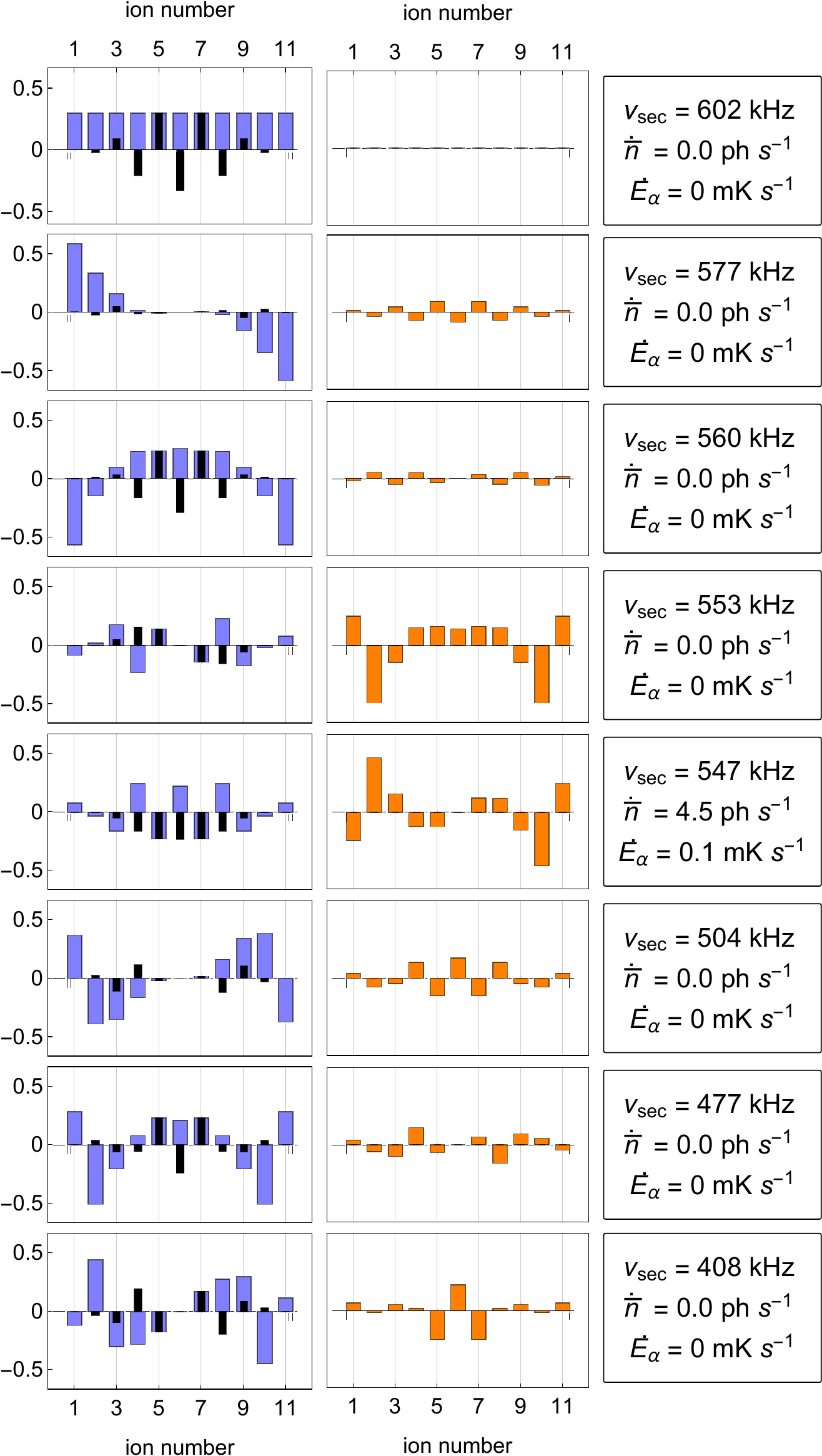}
	\end{center}
	\caption{RF-noise induced heating of a 2D 11-ion crystal (continued).}{
		
	}
\label{fig:C2}
\end{figure}
\begin{figure}[h]
	\begin{center}

		\includegraphics[clip,width=0.8\textwidth]{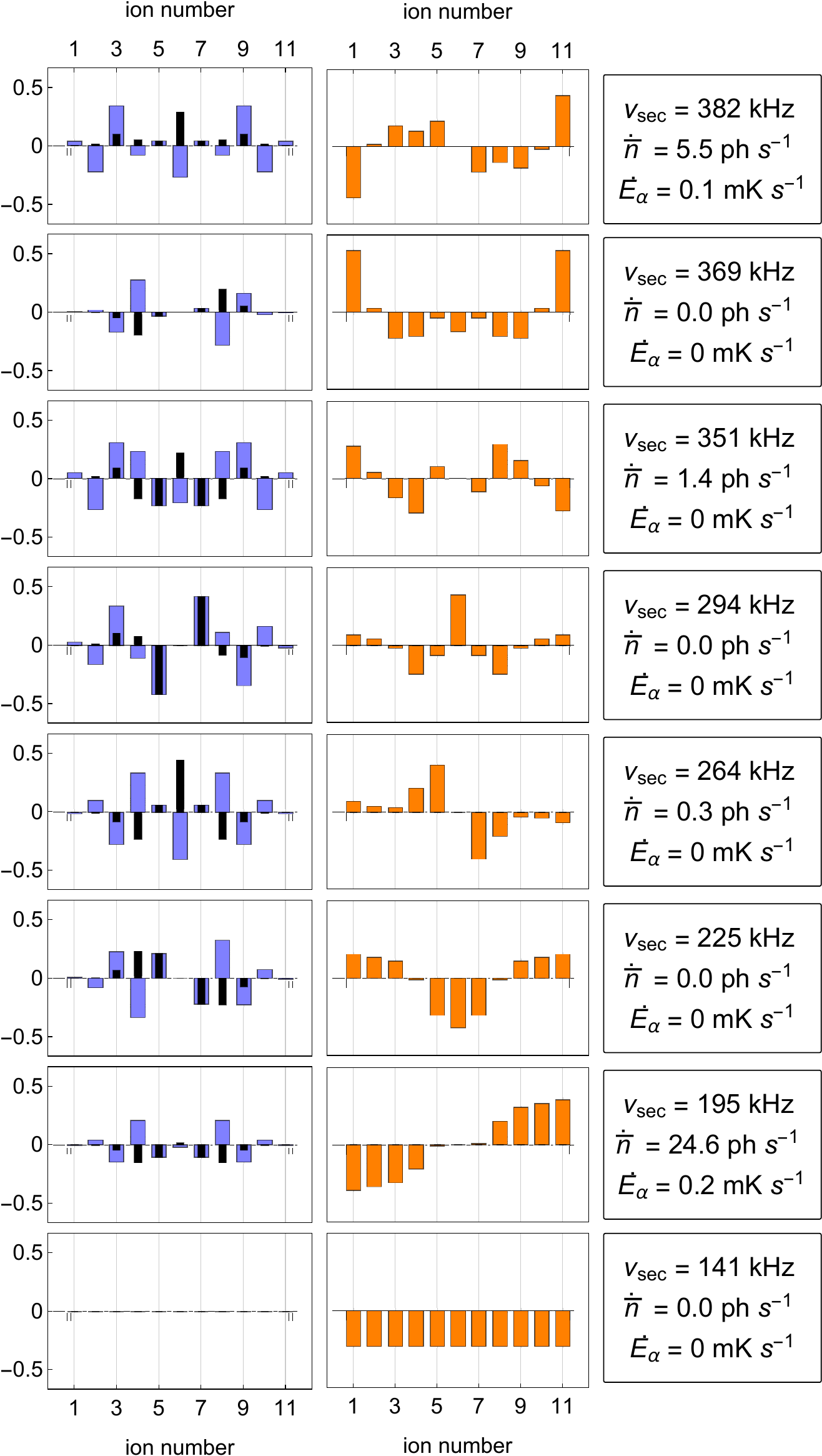}
	\end{center}
	\caption{RF-noise induced heating of a 2D 11-ion crystal (continued).}{
		
	}
\label{fig:C3}
\end{figure}
\FloatBarrier
\section*{References}	

\end{document}